\documentclass{aa}  

\usepackage{placeins}
\usepackage{graphicx}
\usepackage{color}
\usepackage{comment}
\usepackage{hyperref}

\usepackage{txfonts}
\makeatletter
\renewcommand*\aa@pageof{, page \thepage{} of \pageref*{LastPage}}
\makeatother

\begin{document}

\title{Tracking optical variability and outflows across the accretion states of the black hole transient MAXI~J1820+070}
\titlerunning{Optical Minutes-Timescale Variability in MAXI J1820+070}

   \author{M. C. Baglio
          \inst{1}
          \and
          K. Alabarta  \inst{1, 2}
          \and
          D. M. Russell \inst{2}
          \and
          N. Masetti \inst{3, 4}
          \and
          M. M. Messa \inst{5, 1}
          \and
          T. Mu\~noz Darias \inst{6, 7}
          \and
          F. M. Vincentelli \inst{8}
          \and
          S. K. Rout \inst{2}
          \and
          P. Saikia \inst{2}
          \and
          A. Gabuya\inst{9, 10}
          \and
          V. Chavushyan\inst{11}
          \and
          T. Al Qaissieh\inst{9}
          \and
          A. Palado\inst{9}
          }

   \institute{INAF, Osservatorio Astronomico di Brera, Via E. Bianchi 46, I-23807 Merate (LC), Italy\\
              \email{maria.baglio@inaf.it}
         \and
             Center for Astrophysics and Space Science (CASS), New York University Abu Dhabi, PO Box 129188, Abu Dhabi, UAE
        \and
             INAF--Osservatorio di Astrofisica e Scienza dello Spazio, Via Piero Gobetti 101, I-40129 Bologna, Italy
        \and
            Universidad Andr\'es Bello, Av. Fern\'andez Concha 700, Las Condes, Santiago, Chile         
            \and       
            Dipartimento di Fisica, Università degli Studi di Milano, Via Celoria 16, I-20133 Milan, Italy       
            \and
            Instituto de Astrof\'{i}sica de Canarias (IAC), V\'{i}a L\'{a}ctea s/n, La Laguna E-38205, S/C de Tenerife, Spain
            \and
            Departamento de Astrof\'{i}sica, Universidad de La Laguna, La Laguna E-38205, S/C de Tenerife, Spain
            \and
            INAF Istituto di Astrofisica e Planetologia Spaziali, Via del Fosso del Cavaliere 100, I-00133 Rome, Italy
            \and
            Al Sadeem Observatory, Al Wathba South, Abu Dhabi, UAE
            \and
            Rizal Technological University, Mandaluyong City, Philippines
            \and
            Instituto Nacional de Astrofísica, Óptica y Electrónica (INAOE), Luis Enrique Erro \#1, Tonantzintla, Puebla, México, C.P. 72840
            }
\authorrunning{M. C. Baglio et al.}             

   \date{Received August 1, 2025; accepted October 17, 2025}

 
  \abstract{
We present a study of the evolution of the minute-timescale optical variability and of the optical spectroscopic signatures of outflows in the black hole X-ray binary MAXI J1820+070 during its main 2018 outburst and subsequent re-brightenings. Multi-filter, minute-cadence optical light curves were obtained with the Las Cumbres Observatory network and the Al Sadeem Observatory (UAE) over 2018–2020, complemented by archival X-ray data from \textit{Swift}/BAT and XRT and from MAXI. We also acquired contemporaneous low-resolution optical spectra with the 2.1 m OAN San Pedro Mártir (México), the 2.1 m OAGH Cananea (México), and the `G.D. Cassini' 1.5 m telescope at Loiano (Italy).

The optical fractional rms is highest in the hard state and is best described by short time-scale flickering that is stronger at longer wavelengths. This suggests that the minute-timescale optical variability is driven by the jet in the hard state. In this scenario, the variability could be due to variations in the inflow that inject velocity fluctuations at the base of the jet (internal shock model). The variability is then quenched in the soft state, with any residual signal likely associated with variability in the accretion flow. This is in agreement with the accretion-ejection coupling in black hole binaries and confirms that the variability signature of the jet is present at optical wavelengths in all hard states. In the dimmest hard states, any residual optical variability may instead be linked to cyclo-synchrotron emission from the hot flow.

The optical spectra reveal double-peaked emission lines and possible signatures of cold winds during the hard state. Such winds were previously reported in this source during the main 2018 outburst; here we provide one of the first tentative detections of their presence also during the subsequent re-brightenings. The absence of optical wind signatures in the soft state likely reflects a higher level of disc ionisation driven by the increased X-ray flux, which suppresses the low-ionisation features detectable in the optical band.}
  
   \keywords{stars:black holes, stars:jets, stars:winds, outflows
               }

   \maketitle
%

\section{Introduction}
\label{sec:intro}

Black hole low mass X-ray binaries (BH-LMXBs) are compact systems hosting a stellar mass black hole (BH) and a main sequence star with a typical mass $\le$1 $M_{\odot}$.
BHs in these systems undergo a process of accretion of matter and angular momentum from the companion star through Roche lobe overflow, which leads to the formation of an accretion disc around the compact object. BH-LMXBs typically have a transient nature, alternating long (months-years) quiescent states, with low level activity and X-ray luminosities ($10^{31-33}\, \rm  erg\,s^{-1}$), and states of enhanced activity called outbursts, triggered by ionization of hydrogen, which typically last shorter (weeks-months) and display X-ray luminosities $L_{\rm X}=10^{36-38}\, \rm  erg\,s^{-1}$.

At the beginning of an outburst, the source reaches $L_{\rm X}\ge 10^{38} \, \rm erg\,s^{-1}$ very rapidly. For most
of this range in luminosity, the X-ray state is observed to be ``hard'', characterised by high X-ray variability (up to $40\%$ fractional rms in the case of GX 339-4, \citealt{Motta2015}; see also \citealt{Munoz2011}) and
by an X-ray spectrum peaking at $\sim 100$ keV, possibly originating in thermal Comptonisation in
a hot plasma (the corona) around the accretion disc. After a while, the system transitions to the ``soft'' state, and the X-ray spectrum is dominated by a black-body component that peaks at $\sim 1$ keV, probably originating in the accretion disc. In the soft state, variability drops significantly. The system then typically fades slowly, and then transitions back to the hard state before going back to quiescence (see \citealt{Belloni2016} for a review). 

Variability is also expected at lower energies, though the picture is rather complex. Historically, the optical emission during outbursts has been known to be dominated by the accretion disc. In particular, two primary components contribute significantly: the viscously-heated disc, which shows slower variability on timescales of days to weeks, driven by changes in the mass accretion rate from the companion star and by structural evolution in the disc itself (e.g., variations in the temperature profile, or the presence of heating/cooling fronts), and X-ray reprocessing, where variable X-ray emission from the inner disc irradiates the outer disc and/or the companion star surface, producing optical and infrared (OIR) emission that can follow the X-ray light curve with a characteristic smearing due to the finite light travel time and reprocessing efficiency (see, e.g., echo-mapping studies such as \citealt{OBrien2002}, \citealt{VanParadijs1994}).

In addition to these dominant sources, other forms of variability contribute on intermediate timescales (hours to days). These include superhumps produced by precessing or warped accretion discs (e.g., \citealt{Thomas2022}), and ellipsoidal modulations of the companion star, though the latter are typically only visible in quiescence. Furthermore, rapid variability on timescales from minutes down to sub-seconds has been observed in some systems, particularly in the redder bands. This has been interpreted as arising from several possible components: a hot, magnetized, geometrically thick and optically thin inner flow emitting synchrotron radiation \citep{Veledina2013}, wind-related variability (e.g., changes in the blue component of P-Cygni absorption lines; \citealt{Vincentelli2025}), or from the jet, which will be discussed in more detail in the following section.
The combination and relative contributions of these components can lead to diverse OIR variability patterns across a broad range of timescales.

Several studies of BH-LMXBs have shown that part of the optical emission may originate in synchrotron emission from relativistic compact jets \citep[e.g.][]{Corbel02, Chaty2003, Buxton2004, Homan2005, Russell2010, Russell2011, Baglio2018, Saikia2019}. Jets are launched from the central regions of BH-LMXBs and typically produce a flat (optically thick) radio spectrum that transitions to optically thin at OIR frequencies, resulting in a negative power law \citep{Corbel02, Gandhi11, Russell2013a}. 
Interestingly, a strong coupling between accretion and ejection has been observed in several BH-LMXBs, with the emission of compact jets when the outburst is in the hard state, that is then replaced by optically thin discrete ejecta during the hard/soft transitions (visualised as powerful radio flares and resolved, superluminal ejecta seen in radio images). The jet is finally completely quenched at all wavelengths during soft states \citep[for a review, see][]{FenderGallo2014}. In these states, hot winds are often detected through X-ray spectroscopy as blue-shifted absorption lines produced by highly ionised material \citep{Ponti2012, DiazTrigo2016, Neilsen2023}. In addition, recent findings show signatures of the emission of cold winds through optical spectroscopy, particularly with the observation of P-Cyg line profiles in He and H lines in the spectra of some BH-LMXBs \citep[V404 Cyg, V4641 Sgr, MAXI J1820+070 and up to ten other sources;][]{MunozDarias2016, MunozDarias2018, MunozDarias2019, Sanchez2020, Panizo2022}. These cold winds have also been observed during hard states, in some cases simultaneously with radio jets, indicating that both phenomena can coexist depending on the accretion flow configuration.

Jet emission has been found to be strongly variable on short (from sub-second to minute) timescales. Several coordinated X-ray/OIR observations have also shown correlated short-timescale variability with lags of the order of fractions of seconds, with the OIR emission lagging the X-rays ($\sim0.1$ s for several BH LMXBs; \citealt{Gandhi2008}; \citealt{Casella2010}; \citealt{Gandhi2017}; \citealt{Paice2019}). 
This lagged, correlated short-timescale variability  has been explained, e.g., by the internal shock model \citep{Malzac2013, Malzac2014, Malzac2018}, which connects the variations in the accretion flow, observed at X-ray frequencies, to the injection of velocity fluctuations at the jet base. These fluctuations drive internal shocks at large distances from the BH, giving rise to variable synchrotron emission, therefore affecting the overall radio–OIR emission.
This model has been successful so far to reproduce the observations of GX 339-4 \citep{Drappeau2017, Malzac2018, Vincentelli2019}, of the neutron star (NS) X-ray binary 4U 0614+091 \citep{Marino2020}, and has been also suggested to explain the high fractional rms observed in the OIR for the BH-LMXBs MAXI J1535-571 \citep{Baglio2018, Vincentelli2021}, MAXI J1836-194 \citep{Peault2019} and MAXI J1820+070 \citep{Tetarenko2021}.

\section{MAXI J1820+070}

The optical transient ASASSN-18ey was first detected on 2018 March 6 by the ASAS-SN survey (V=14.9; \citealt{ATel11400}). A few days later, on March 11, the source was independently identified in X-rays by the MAXI/GSC nova alert system, which reported it as a bright, uncatalogued X-ray source and designated it MAXI J1820+070 \citep{Atel11399}. 
Follow-up on March 13 with the Las Cumbres Observatory (LCO) network 1-m telescopes revealed a brightened optical counterpart, several magnitudes above its Pan-STARRS quiescent level. The strong X-ray emission and its location on the optical/X-ray luminosity plane indicated a low-mass X-ray binary (LMXB) with a likely black hole primary \citep{ATel11418}, later confirmed by dynamical measurements showing a mass ratio of $q \sim 0.04$ and a compact object mass exceeding the neutron star limit \citep{torres2019, Torres2020}. VLBI parallax placed the source at $2.96 \pm 0.33$ kpc, constraining its luminosity and jet energetics \citep{Atri2020}.

Short-timescale optical variability was observed, including $>100\%$ amplitude flares on $\sim$100 ms and $\sim$10 ms scales \citep{Atel11426}. Sub-second variability, strongest in red, was detected with ULTRACAM on the NTT, suggesting a jet origin \citep{Atel11437}. A flat radio spectrum observed on March 18 with RATAN-600 supported this interpretation \citep{Atel11439}, and later studies linked the fast optical variability to synchrotron emission from the jet \citep{Paice2019, Tetarenko2021, Thomas2022}. In the near infrared (NIR), $Ks \sim 10.2$ with $\sim25\%$ rms variability was detected on March 19 using VLT/HAWK-I \citep{Atel11451}. On April 8–9, a bright mid-IR counterpart ($\sim$0.3 Jy) was found, making MAXI J1820+070 the brightest mid-IR transient LMXB observed \citep{Russell2018ATel11533, Echiburu2024}.

Following an initial rise and hard-state plateau, \textit{NICER} observed a flux decline in mid-May 2018, followed by an increase with spectral softening, indicating a transition to the hard-intermediate state by early July and then to the soft state \citep{Atel11820}. \textit{AMI-LA} detected quenched radio emission and flaring, consistent with this transition \citep{Atel11827}. The evolution has been detailed through multiwavelength studies \citep{Bright2020, Homan2020} and X-ray reverberation lag studies \citep{Kara2019}. Broadband X-ray data from \textit{Insight-HXMT} revealed an extended, jet-like corona in the hard state \citep{You2021}, though recent X-ray polarisation results on different systems (e.g. Swift J1727) suggest that the corona is generally unlikely to be extended in this way in the hard state. \citet{Espinasse2020} reported X-ray sources aligned with the radio jets, best interpreted as discrete ejecta interacting with the ISM on large scales, rather than with the corona or the hard state compact jets.

A sharp rise in hard X-ray flux after September 22, 2018 (MJD 58383), detected by MAXI/GSC, signaled a soft-to-hard transition \citep{Atel12057}. Radio detection by AMI-LA on September 24 confirmed the reactivation of the jet \citep{Atel12061}, with optical signs of jet activity (increased flux and redder color) observed by LCO on October 12 \citep{BaglioATel12128, Echiburu2024, OzbeyArabaci2022}. After nearing quiescence \citep{Atel12534}, the source showed an unexpected optical re-brightening \citep{Atel12567,Atel12596}.

The re-brightening was also seen in X-rays ($Swift$/XRT, March 10, 2019; \citealt{Atel12573}) and radio (AMI-LA; \citealt{Atel12577}). Flux declined through April–May, with dimming in both X-rays and optical \citep{Atel12732, Atel12747}. A second brightening followed in August 2019, first in the optical \citep{Atel13014}, then in X-rays and radio \citep{Atel13025, ATel13041}, peaking in 10 days and lasting ~70 days. A third re-brightening occurred in February 2020 in both optical and X-rays \citep{Atel13502, Atel13530}. By June 29, 2020, the source reached $g' = 18.79 \pm 0.01$, still ~0.6 mag above its Pan-STARRS quiescent level. Another optical brightening was seen in March 2021 by LCO/XB-NEWS \citep{Baglio2021a}, with flux declining by late April but remaining above quiescence \citep{Baglio2021b}. Low-level activity continued until June 2023, when the source finally returned to full optical quiescence \citep{ATel16192}.

\section{Short timescale variability: Observations and data analysis}

In this paper we report on an optical campaign performed on MAXI J1820+070. The system was regularly monitored since its first activity with the LCO network of robotic 1-m and 2-m telescopes, using the optical filters $Y$, $i'$,$r'$, $R$, $V$, $g'$. 
In addition to these, many observations have been collected with the Meade LX850 16-inch (41-cm) telescope of the Al Sadeem Observatory (UAE), as part of the monitoring that started in 2018, using the 'blue', 'green' and 'red' Baader LRGB CCD-Filters (that approximately correspond to the central wavelengths of the g', V and R filters, respectively; see \citealt{Russell2018ATel11533} and \citealt{Russell2019} for details). Although the full optical light curves are shown in the upper panel of Fig. \ref{lc_fig} (until MJD 59000; 31 May 2020), in this work we will specifically focus on short-timescale (minutes) observations of the target, that happened sporadically in the period of observations with the $i'$ and $g'$ filters with LCO (Table \ref{log_LCO}) and with all the Al Sadeem Observatory filters (Table \ref{log_AlSadeem}). 


We also retrieved public X-ray observations taken with \textit{Swift}/BAT, \textit{Swift}/XRT and MAXI in order to have simultaneous X-ray observations concurrent with the optical monitoring.

\subsection{Optical photometry with LCO}

MAXI J1820+070 was regularly monitored since its discovery with the LCO 1-m and 2-m telescopes, as part of an ongoing monitoring of $\sim 50$ LMXBs, coordinated by the Faulkes Telescope Project\footnote{\url{http://www.faulkes-telescope.com}} \citep{Lewis2008}. The exposure time varied depending on the filter, the phase of the outburst, and the aim of the observation. For general outburst monitoring, the exposure time varied between 20s and 200s in all filters; for the timing observations instead, the single exposures lasted in general 20-60s, with filters alternating in most cases. In Table \ref{log_LCO}, a complete log of the fast-timing observations is reported for completeness.

Magnitudes were extracted using multi-aperture photometry (MAP; \citealt{Stetson90}) performed with the "X-ray Binary New Early Warning System" (XB-NEWS) pipeline \citep{Russell2019, Goodwin2020}. For each reduced image, XB-NEWS first detects sources using SExtractor \citep{Bertin1996} and solves for the astrometric solution using astrometry.net \citep{Lang2010}, matching against Gaia DR2 positions. If the target is not detected within 1'' of its known coordinates at the default threshold, the pipeline re-runs detection at a lower threshold. If still undetected, forced photometry is performed at the target position.
Photometry is carried out for all sources using both MAP and fixed-aperture photometry, with aperture radii scaled to an appropriate multiple of the PSF FWHM. Light curves are then constructed using the DBSCAN clustering algorithm \citep{Ester96}.
Calibration of $i'$ and $g'$-band magnitudes is performed against an enhanced version of of the ATLAS-REFCAT2 catalogue \citep{Tonry2018}, which incorporates Pan-STARRS DR1 \citep{Chambers2016} and APASS DR10 \citep{Henden2018}. A photometric model is fit to the light curves (in magnitudes) of matched and unmatched sources, including spatially variable zeropoints, PSF-based terms, and source-specific mean magnitudes \citep{Bramich2012}. The model excludes color terms due to limited multi-band coverage, introducing only small systematic errors ($<(1-2)\%$) in absolute calibration. The model is fitted iteratively with outlier down-weighting to mitigate the effect of intrinsic variability.
For SDSS $i'$ and $g'$ filters, XB-NEWS uses the Pan-STARRS1 standard $i_{\rm P1}$ and $g_{\rm P1}$ magnitudes in the AB system. The final model is applied to calibrate all light curves, including that of the target.
If forced photometry was required for the target, measurements with uncertainties >0.25 mag were discarded as unreliable. For more details about the pipeline concept and implementation, see \citep{Russell2019, Goodwin2020}.

\begin{center}
\begin{figure*}
\includegraphics[scale=0.83]{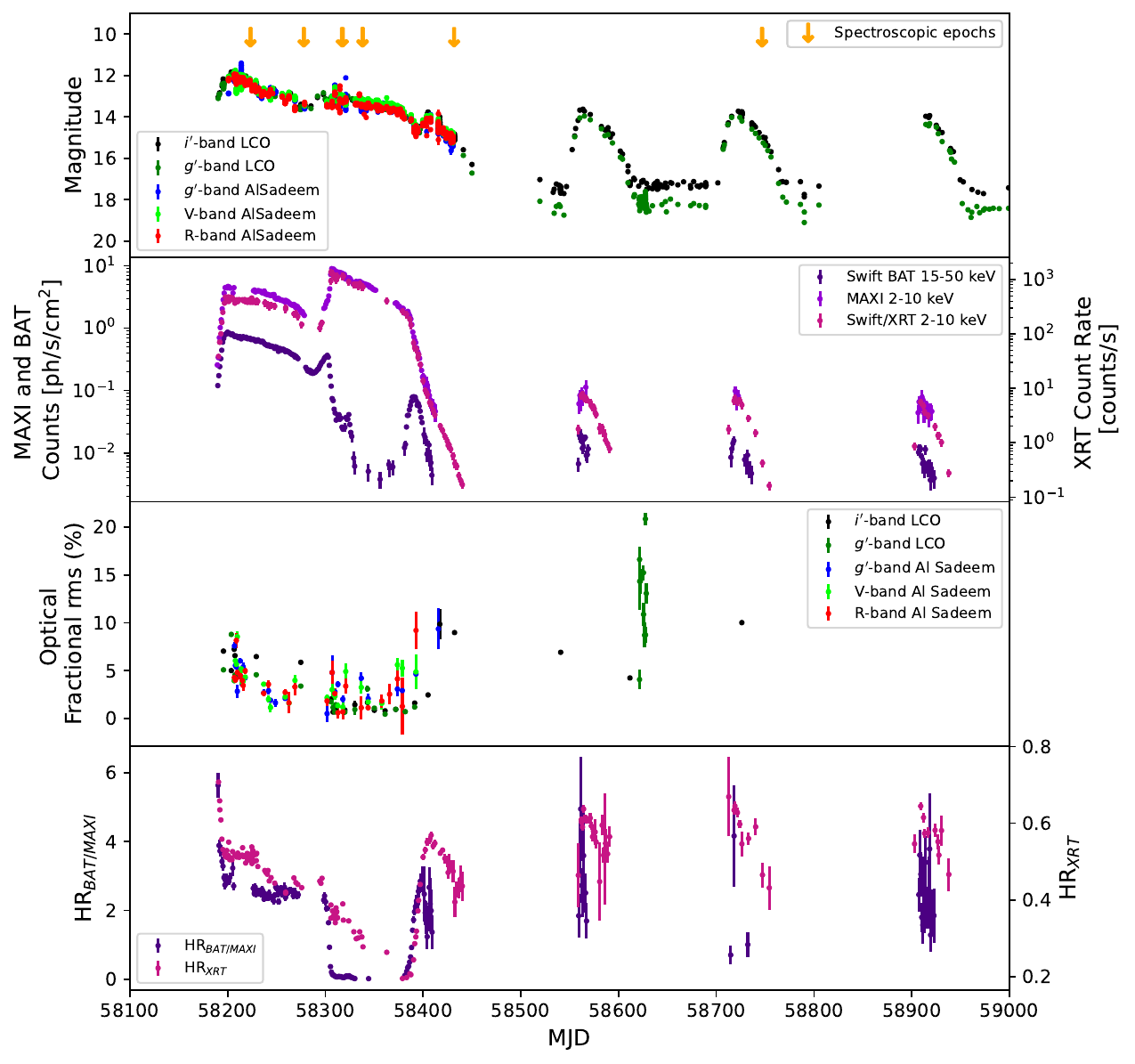}
\caption{\textit{First panel}: Optical LCO ($g'$ and $i'$) and Al Sadeem ($g'$, $V$, $R$) light curves of MAXI J1820+070 during its 2018/2019 outburst. Magnitudes are not corrected for reddening. \textit{Second panel}: \textit{Swift}/BAT (purple), \textit{Swift}/XRT (pink) and MAXI/GSC (dark purple) count rates vs. MJD. \textit{Swift}/BAT counts have been divided by 100 for graphic purpose. \textit{Third panel}: MJD vs. fractional root mean square (rms) variability amplitude evaluated for the time-resoved optical light curves obtained with the LCO and Al Sadeem observations.\textit{Fourth panel}: MJD vs. Hardness ratio evaluated as BAT/MAXI count rates ($\rm HR_{\rm BAT/MAXI}$) and XRT (2-10 keV)/XRT (0.5-2 keV) count rates ($\rm HR_{\rm XRT}$).}
\label{lc_fig}
\end{figure*}
\end{center}

The resulting $g'$ and $i'$ light curves for the monitoring during the 2018 outburst and the three subsequent reflares (until MJD 59000; 2020 May 31) are shown in the upper panel of Fig. \ref{lc_fig}. Part of these data are also pulished in \citet{Russell2019, Echiburu2024, banerjee2024, Yang2025, Bright2025}. For single light curves, see Appendix \ref{lco_appendix}.

\subsection{Optical photometry with the Al Sadeem Observatory}

MAXI J1820+070 was observed with the Meade LX850 16-inch (41-cm) telescope of the Al Sadeem Observatory (UAE)\footnote{\url{https://alsadeemastronomy.ae}}, using the `blue', `green' and `red' Baader LRGB CCD-Filters (central wavelengths 463, 538, 638 nm, respectively), that approximately correspond to the central wavelengths of the g' (475 nm), V (545 nm) and R (641 nm) filters, respectively. Observations were performed during most of the source activity, with exposure times ranging from 30 to 60s. Photometric measurements on some dates were previously reported in \cite{BaglioATel12128,Atel12596,Baglio2021b,Russell2018ATel11533,Russell2019,Echiburu2024}. Similar to the LCO observations, on several dates sequences of observations were acquired in the three filters, in order to measure short-timescale variability. Here we report on observations acquired during the main outburst, between MJD 58200 (2018 March 23) and 58430 (2018 November 08), including rms variability measurements for the first time. A complete log of the variability observations is presented in Table \ref{log_AlSadeem}.

All images were bias and flat-field corrected using standard procedures and aperture photometry was performed using the {\tt daophot} tool within MIDAS\footnote{MIDAS (Munich 
Image Data Analysis System) is developed, distributed and maintained 
by ESO and is available at {\tt http://www.eso.org/projects/esomidas/}}.
Calibration was performed against a group of seven isolated field stars, with magnitudes calibrated in the APASS catalogue in the $g'$, $V$ and $r'$ filters. Since the Al Sadeem 'red' filter corresponds to the Bessell $R$ filter and not the SDSS $r'$ filter, while the APASS catalogue provides magnitudes in $r'$, we applied the appropriate transformation from $r'$ to $R$ using the color equations provided by \citet{Jordi2006}.
The results of the photometry are shown in Fig. \ref{lc_fig}, upper panel.




\subsection{X-ray campaign}\label{sec:x-rays}

MAXI J1820+070 was monitored with different X-ray satellites since it was first detected by MAXI/GSC on 2018 March 11 (MJD 58188). For this work, we consider the public observations from the monitoring of the 2018/2019 outburst by \textit{Swift}/BAT (15-150 keV)\footnote{\url{https://swift.gsfc.nasa.gov/results/transients/}} and \textit{Swift}/XRT (0.5-10 keV)\footnote{\url{https://www.swift.ac.uk/swift_portal/}}, and by MAXI/GSC (2-10 keV) \footnote{\url{http://maxi.riken.jp/top/index.html}}.
 We converted the fluxes from both \textit{Swift}/BAT and MAXI into Crab units to normalise their instrumental responses and thus derive a hardness ratio between them. For this, we used the following conversion factors: 1 Crab$=$0.22~$\rm ph/s/cm^2$ and 1 Crab$=$3.45~ $\rm ph/s/cm^2$ for \textit{Swift}/BAT and MAXI, respectively \footnote{\url{http://maxi.riken.jp/home/asai/maxibatHR/}}. Then, we defined the hardness ratio $\rm HR_{\rm BAT/MAXI}$ as the ratio between the flux of \textit{Swift}/BAT and MAXI in Crab units (Fig.\ref{lc_fig}, last panel).
Similarly, using \textit{Swift}/XRT data we evaluated the hardness ratio $\rm HR_{\rm XRT}$, defined as the ratio of the 2$-$10 keV XRT count rates to the 0.5$-$2 keV XRT count rates (Fig.\ref{lc_fig}, last panel).

\section{Short timescale variability: Results}\label{sec:results}

\subsection{Optical variability and X-ray hardness analysis}\label{sec:optrms_hr}

To estimate the amount of short timescale variability of the optical light curves, we evaluated the fractional root mean square (rms) variability amplitude following \cite{Vaughan2003}. In particular, the quantity is defined as follows:

\begin{equation}
    \rm rms (\%) = \frac{\sqrt{S^2-\overline{\sigma^2_{\rm err}}}}{\overline{x}}\,, 
\end{equation}

where $\sigma_{\rm err}$ is the error on the single flux value, that is then squared and averaged for each sample, and $\overline{x}$ is the mean magnitude value of the sample. S, instead, is the variance of the sample, and it is defined as:

\begin{equation}
    S^2=\frac{1}{N-1}\sum_{i=1}^{N} ( x_i - \overline{x})^2\, ,
\end{equation}

where $N$ is the total number of observations in the sample, and $x_i$ is the single flux value of the $i$-th observation in the sample.

The fractional rms variability amplitude is found to vary from epoch to epoch, in all bands, as can be observed in Fig. \ref{lc_fig} (third panel), going from $\sim 0$ to $\sim 20\%$. This variability is sampled over timescales corresponding to frequencies approximately between $\sim 7 \times 10^{-4}$ and $\sim 0.03,\mathrm{Hz}$, depending on the duration and cadence of the individual light curves.

During the main outburst (March 2018 - February 2019), the fractional rms variability amplitude calculated for both the LCO and Al Sadeem datasets is found to be higher ($\sim 5\%$ and $\sim 6\%$ on average in the LCO $g'$ and $i'$ band, respectively) during the long hard-state plateau happening before $\sim$ MJD 58300 (2018 July 1), and then gets lower (on average, $\sim 1\%$ in both bands) after the transition to the soft-intermediate state, until MJD 58383 (2018 September 22, i.e. the time of the transition to the hard X-ray spectral state; see Fig. \ref{lc_fig}). 
After that, the fractional rms variability amplitude increases again, reaching $\sim 9\%$ in $i'$-band.
During the first and second re-brightenings, only one $i'$-band fractional rms variability amplitude measurement is available for each, with values of $\sim 7\%$ and $\sim 10\%$, respectively.
An interesting feature is observed on MJD $\sim$ 58620-30 (i.e. 2019 May 17-27), with the fractional rms variability amplitude in the $g'$ band suddenly increasing, reaching up to $\sim20\%$ on MJD 58627.5. At that time, the source was in an apparent quiet state soon after the end of the first re-brightening, with a magnitude of $g'\sim 16$. 

\begin{center}
\begin{figure}
\includegraphics[scale=0.6]{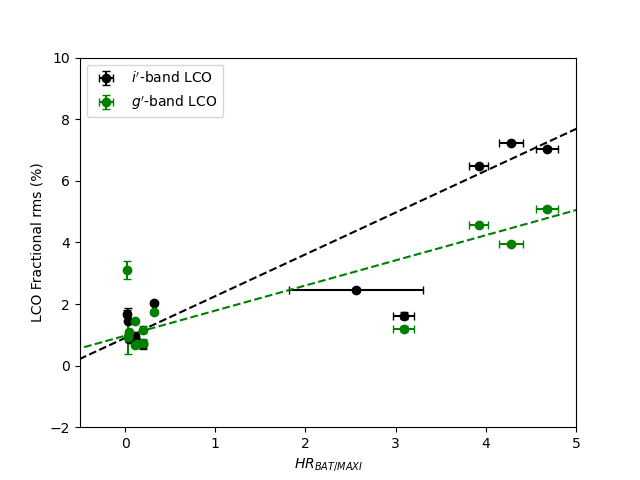}
\caption{Hardness ratio (calculated as BAT/MAXI) vs. fractional rms of the optical LCO $g'$ (green dots) and $i'$ (black dots) data. Linear fits to the data are also represented with dashed green and black lines, respectively, to show the increasing trend of the fractional rms with the hardness. 
}
\label{Fig:rms_LCO}
\end{figure}
\end{center}

The last two panels of Fig. \ref{lc_fig} suggest a possible correlation between the fractional rms variability amplitude of the optical curves and the hardness evolution of the X-ray spectrum.
Motivated by this, we investigated whether changes in the optical fractional rms could be directly linked to variations in the X-ray hardness. 
The results are shown in Fig. \ref{Fig:rms_LCO} and \ref{Fig:rms_AlSadeem} (top panel) for the LCO and Al Sadeem datasets, respectively, using as hardness ratio the quantity $\rm HR_{\rm BAT/MAXI}$.
A clear correlation is observed in the LCO dataset, with the optical rms increasing toward harder spectra across all filters. The Al Sadeem data show greater scatter, likely due to lower data quality, and exhibit only a mild correlation, most noticeable in the $R$-band. The correlation parameters are listed in Table~\ref{tab:correlations}.

\begin{table}
\caption{Results of the correlation study using the hardness ratio calculated as the ratio between the BAT (hard) and the MAXI (soft) X-ray fluxes vs. optical fractional rms of the LCO $g'$ and $i'$ and Al Sadeem $g'$, $V$ and $R$ lightcurves.}            
\label{tab:correlations}      
\centering                       
\begin{tabular}{c |c c c}       
\hline    
\multicolumn{4}{c}{HR $BAT/MAXI$}\\
\hline   
Filter & Intercept (\%) & Slope (\%)& Pearson coefficient \\
\hline
\multicolumn{4}{|c|}{LCO}\\
\hline
$g'$     & $0.98\pm0.16$ & $0.39\pm 0.07$ &0.81 \\
$i'$     & $0.91 \pm 0.18 $ & $0.64\pm 0.09$ & 0.89\\
\hline
\multicolumn{4}{|c|}{Al Sadeem}\\
\hline
$g'$    & $ 2.73\pm0.19 $ & $ 0.10\pm 0.09 $ & 0.05 \\
$V$     & $ 1.70 \pm 0.19  $ & $0.52 \pm 0.09 $ & 0.16 \\
$R$     & $ 1.39 \pm 0.21  $ & $0.75 \pm 0.11 $ & 0.34 \\
\hline
\hline
\end{tabular}
\end{table}

In all bands, a positive Pearson coefficient indicates a linear correlation between the fractional optical rms variability amplitude and the hardness of the X-ray spectrum. Specifically, Pearson coefficients greater than 0.8 for the LCO $g'$ and $i'$ bands suggest a strong or very strong linear correlation between these two quantities.

\begin{center}
\begin{figure}
\includegraphics[scale=0.6]{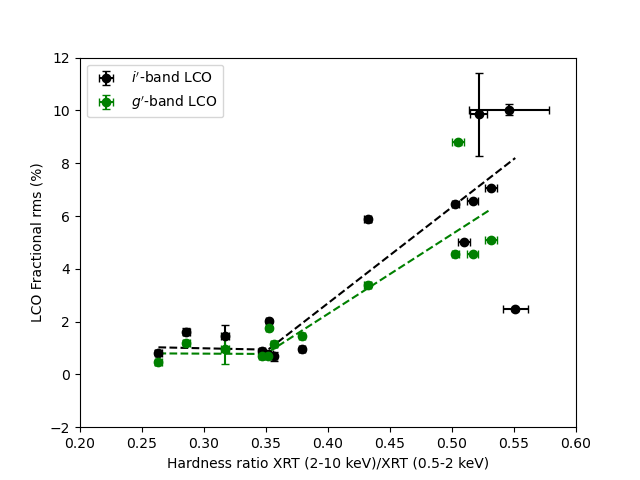}
\caption{Fractional rms vs. Hardness ratio calculated as the XRT 2-10 keV / XRT 0.5-2 keV flux and the optical fluxes obtained with LCO in the $i'$ (black) and $g'$ (green) bands. Broken power law fits are also represented with black and green dashed lines, respectively, to show the increasing trend of the fractional rms with a hardness >0.35. Results of the fits are reported in Tab. \ref{tab:correlations_XRT}.}
\label{fig:rms_LCO_XRT}
\end{figure}
\end{center}

Similarly, for both the LCO and Al Sadeem datasets we plotted the optical fractional rms variability amplitude against the hardness ratio $\rm HR_{\rm XRT}$
This correlation analysis employs softer X-ray photons than those used in Fig. \ref{Fig:rms_LCO} and \ref{Fig:rms_AlSadeem} (top panel), allowing us to probe the rms-hardness correlation for X-ray photons emitted by softer components, such as the accretion disc. The results of this analysis are shown in Fig. \ref{fig:rms_LCO_XRT} and \ref{Fig:rms_AlSadeem} (bottom panel).

The trend of the optical fractional rms variability amplitude with $\rm HR_{\rm XRT}$ does not follow a simple positive correlation;
therefore, we fit the data using a broken-line model. In the LCO dataset, the fit reveals that, for low values of $\rm HR_{\rm XRT}$, the slope is close to zero in the $i'$ and $g'$ bands. After the break, occurring at $\rm HR_{\rm XRT} \sim 0.35$, the data exhibit a strong positive correlation, with Pearson coefficients of 0.77 and 0.84 in the two bands, respectively.  We note that $\rm HR_{\rm XRT} \lesssim 0.35$ correspond to MJD $\sim 58310-58380$, which, according to the classification by \citet{Shidatsu2019} and \citet{Fabian2020}, marks the soft state of the main outburst.
Similar trends are seen in the Al Sadeem data, though the break is less well defined due to lower data quality and higher scatter. Fit results are summarized in Table \ref{tab:correlations_XRT}.


\begin{table*}
\caption{Results of the correlation study using HR$_{\rm XRT}$ vs. optical fractional rms of the LCO $g'$ and $i'$ lightcurves (Fig. \ref{fig:rms_LCO_XRT}) and the Al Sadeem $g'$, $V$ and $R$ lightcurves (\ref{Fig:rms_AlSadeem}, bottom panel).}            
\label{tab:correlations_XRT}      
\centering                       
\begin{tabular}{c |c c c c c}       
\hline    
\multicolumn{6}{c}{$\rm HR_{\rm XRT}$}\\
\hline   
Filter & Slope 1 (\%)& Break HR & Break rms (\%) & Slope 2 (\%)& Pearson coefficient \\
\hline
\multicolumn{6}{|c|}{LCO}\\
\hline
$g'$   & $-0.23\pm8.73$ & $0.35\pm 0.02$ &$0.77\pm0.33$ & $30.18\pm5.68$ & 0.84 \\
$i'$    & $-1.02\pm8.80$ & $0.35\pm 0.02$ &$0.93\pm0.39$ & $36.41\pm5.61$ & 0.77 \\
\hline
\multicolumn{6}{|c|}{Al Sadeem}\\
\hline
$g'$   & $-8.74\pm2.37$ & $0.45\pm 0.01$ &$1.99\pm1.40$ & $47.47\pm4.08$ & 0.37 \\
$V$    & $-23.30\pm5.61$ & $0.39\pm 0.01$ &$1.05\pm2.89$ & $30.33\pm1.96$ & 0.27 \\
$R$    & $-12.57\pm15.06$ & $0.36\pm 0.02$ &$1.76\pm7.47$ & $15.65\pm2.86$ & 0.33 \\

\hline
\hline
\end{tabular}
\end{table*}

\subsection{X-ray fractional rms amplitude}

\begin{figure*}
\centering
\includegraphics[width=\textwidth]{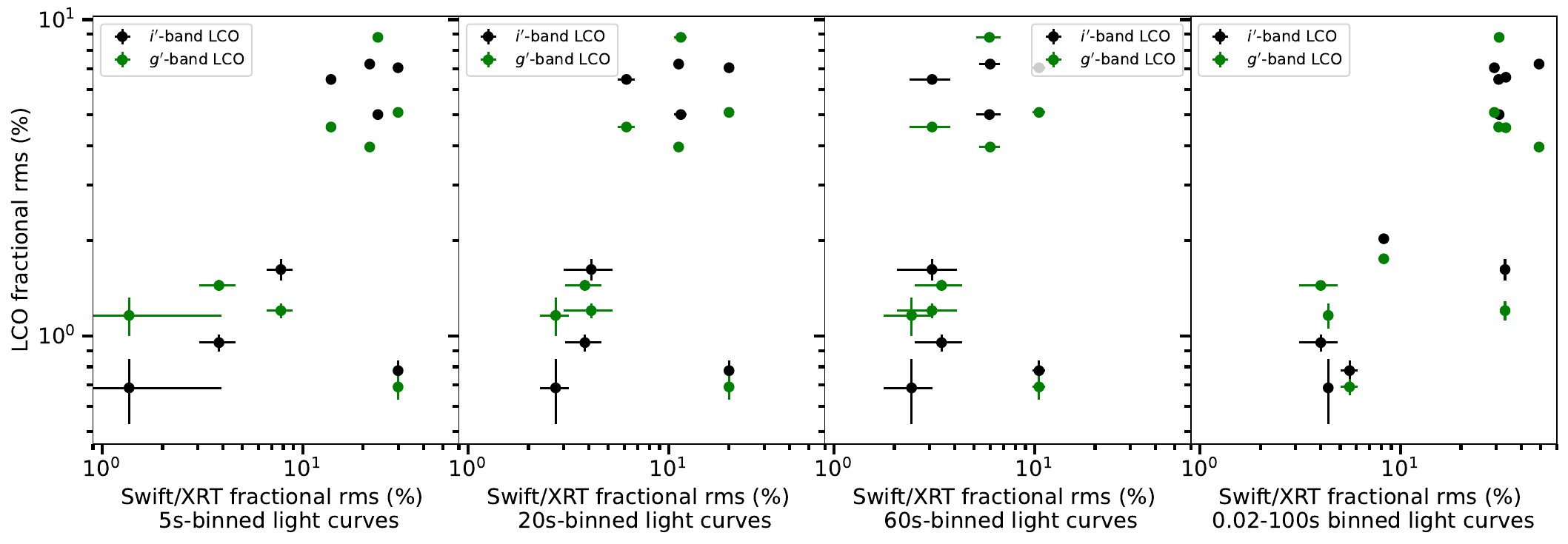}
\caption{Fractional rms of the optical LCO $g'$ (green dots) and $i'$ (black dots) data vs the \textit{Swift}/XRT fractional rms amplitude (\%) in the 2.0$-$10.0 keV energy band. The first three panels show the \textit{Swift}/XRT rms obtained from 5s-, 20s- and 60s-binned light curves, respectively. The rightmost panel shows the \textit{Swift}/XRT broadband-frequency rms 0.01$-$64~Hz (i.e. 0.02-100s) in the x-axis. Horizontal and vertical error bars represent one-sigma uncertainties calculated analytically for each respective method.}
\label{Fig:xray_rms}
\end{figure*}

To reinforce the study of the link between the optical variability and the spectral states, we obtained the fractional rms amplitude in X-rays using \textit{Swift}/XRT observations in the 2$-$10 keV energy band. We used two complementary methods. First, we selected specific time-bins to approximate the nominal exposure times of the LCO optical data. We note, however, that the effective time resolution of the optical light curves is coarser (typically $\sim$150 s) due to filter switching and readout time. From the 5s, 20s and 60s background-subtracted \textit{Swift}/XRT light curves, we calculated the fractional rms values following the same procedure described for the optical data \citep[Sec. \ref{sec:optrms_hr};][]{Vaughan2003}. Second, event-mode \textit{Swift} data were analysed in the Fourier domain: Leahy-normalised power-density spectra (PDS) were computed for each observation segment \citet{Leahy83}, whereas the Poisson-noise contribution was estimated from frequencies in the range of 150$-$281 Hz and subsequently subtracted. The resulting PDS were normalised to rms units following \citet{VanderKlis1995}. Finally, we integrated the variability in the frequency range 0.01$-$64 Hz (corresponding to $\sim 0.02-100$s).   

Figure \ref{Fig:xray_rms} shows the optical fractional rms measured by LCO in the $g^{\prime}$ (green dots) and $i^{\prime}$ (black dots) bands plotted against the X-ray fractional rms amplitudes calculated in the 2.0–10.0 keV band for the three different timescales described above: 5s, 20s, and 60s (first three panels, respectively), as well as for the full $\sim$0.02–100s range (rightmost panel). We note that the latter was obtained by integrating the PDS over the corresponding frequency range, rather than by binning the light curves.

As the timescale decreases, we observe an increasingly positive correlation between the optical and X-ray rms (despite some scatter), with Pearson coefficients of approximately 0.5 (moderate positive correlation), 0.3, and 0.2 (both weak positive correlations) for the three timescales, respectively. In all cases, the $i^{\prime}$-band rms shows a better correlation with the X-ray rms than the $g^{\prime}$-band rms. We note that all plots include a clear outlier corresponding to MJD 58305, which falls during the hard-to-soft transition. During this phase, the X-ray rms remains high, while the optical variability appears significantly reduced. This suggests a change in the dominant source of fast optical variability, potentially linked to evolving emission mechanisms during the state transition.

Finally, the optical rms shows a moderate-to-strong positive correlation with the integrated X-ray rms in the 0.01–64 Hz range, with Pearson coefficients of 0.6 and 0.8 in the $g^{\prime}$ and $i^{\prime}$ bands, respectively. 
This reflects that the integrated rms provides a reliable estimate of the total X-ray variability, as it encompasses fluctuations across all timescales. The observed correlation with the optical rms may suggest that including intermediate-timescale variations results in a better alignment with the optical response.



\section{Short timescale variability: Discussion}

The origin of rapid optical variability in BH-LMXBs, observed on sub-second to minute timescales, is a topic of ongoing debate in the astrophysical community. 
Optical emission in black hole binaries can arise from several components, including X-ray reprocessing in the outer disc (e.g., \citealt{King1998}), synchrotron radiation from a compact jet (e.g., \citealt{Malzac2018}), or from a magnetized hot accretion flow (e.g., \citealt{Fabian1982}; \citealt{Veledina2011}), magnetic loop reconnection in the disc \citep[e.g. ][]{Zurita2003})
These components can drive variability on different timescales and leave distinct observational signatures, which can help disentangle their relative contributions.
Various models have been proposed to explain the complex optical variability observed in BH-LMXBs (e.g. \citealt{Fabian1982}; \citealt{Merloni2000}; \citealt{Casella2009}; \citealt{Veledina2013}; \citealt{Malzac2013}; \citealt{Malzac2014}; \citealt{Uttley2014}; \citealt{Tetarenko2021}), although few can account for all aspects of variability, including cross-correlation, autocorrelation, and frequency/X-ray state dependence. In the following, we will examine some of the possible origins of the variability observed for MAXI J1820+070 during the 2018-2019 outburst.


\begin{figure}
    \centering
    \includegraphics[width=1\linewidth]{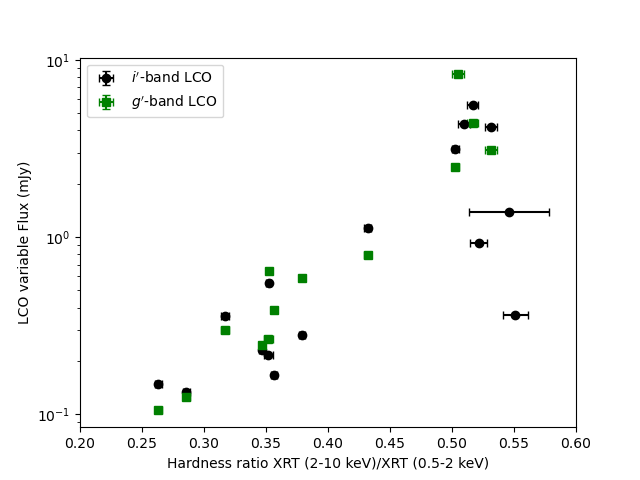}
    \caption{Absolute rms evaluated for LCO $i'$ and $g'$ band observations vs. Hardness of the X-ray spectrum, evaluated using the two \textit{Swift}/XRT energy bands (2-10 for hard; 0.5-2 for soft).}
    \label{fig:absolute_rms}
\end{figure}

\subsection{Thermal reprocessing}

First, we examined whether the observed optical variability could be attributed to X-ray reprocessing. In this scenario, the optical signal represents a delayed and smoothed version of the X-ray variability, shaped by the response of the accretion disc \citep{OBrien2002,vincentelli2020}. In the frequency domain, this leads to a modification of the X-ray variability amplitude depending on the transfer function of the disc \citep{hynes2005, Hynes2006ApJ, veledina2017}. The amplitude of the reprocessed emission is expected to be inversely proportional to the area of its production. Given the large extent of the optically emitting regions of the disc ($\approx$ 1-10 light seconds, see e.g. \citealt{OBrien2002,hynes2005, Hynes2006ApJ}), the reprocessed amplitude is seen to be much lower (more than an order of magnitude in fractional units, i.e. never beyond 10\%) than the driving one (see \citealt{Gandhi2010} for the BH LMXB GX 339-4, or \citealt{vincentelli2020,Vincentelli2023} for the case of NS LMXBs). 


Since rms variability is computed over a specific frequency range, it is important to apply the same frequency interval when comparing optical and X-ray rms values. In our case, the frequency range probed by the optical data is constrained by the time resolution and total duration of the observations. Because most observations alternate between $g'$ and $i'$ filters and include CCD readout times, the effective time resolution is more than twice the exposure time. For example, in the frequent cases in which we have 16 data points per filter over 40 minutes (Tab. \ref{log_LCO}), the optical data probe variability in the frequency range of approximately $4.2 \times 10^{-4}$ to $6.7 \times 10^{-3}$ Hz. This constraint limits our comparison to the third panel of Fig.\ref{Fig:xray_rms}. In that panel, the optical rms is substantially higher than predicted by a reprocessing-only scenario. In some epochs, the optical rms even exceeds the X-ray rms, indicating that additional components beyond thermal reprocessing must contribute to the observed optical variability.
It is also worth noting that quasi-periodic oscillations (QPOs) were reported during the hard state, at both optical and X-ray wavelengths \citep{Paice2021,Mao2022,Fiori2025}. QPO frequencies as low as $\sim 50$ mHz were reported (periods of $\sim 20$ sec), so variability from these QPOs could contribute to the rms variability, although no obvious QPOs are present by inspecting the light curves (Fig.~\ref{fig:appendix2}). The optical QPOs were sharper (lower FWHM) than X-rays QPOs \citep{Mao2022}, so the optical QPOs cannot be due to reprocessing. In addition, longer timescale oscillations were reported on timescales close to the orbital period \citep{Thomas2022,Fiori2025}, which were interpreted as high amplitude superhumps from a warped outer disc. Some of the slow trends seen in our light curves around the hard-to-soft transition and in the hard state could be due to sampling of a small part of these modulations.

We also constructed an absolute rms spectrum by multiplying the fractional rms variability by the average de-reddened flux in the two available bands. The resulting correlation with the X-ray hardness ratio ($\rm HR_{\rm XRT}$) is shown in Fig. \ref{fig:absolute_rms}. Both bands display a strong and comparable correlation with X-ray spectral hardness, apart from a few outliers. These may correspond to hard-state observations at relatively low luminosities, where the fractional rms remains high (typically $5-10\%$), but the absolute flux is low. Given that the hard state extends down to quiescence, such cases are expected, especially since our dataset includes only limited coverage at these lower luminosities. 
The spectrum of the variable optical emission does not exhibit a blue shape, which argues against a thermal origin such as disc reprocessing. Instead, the observed properties are more consistent with a non-thermal origin, such as synchrotron emission.
In BH-LMXBs, synchrotron radiation can originate either from a compact, magnetized hot flow \citep{Veledina2013} or from a collimated relativistic jet \citep{Malzac2014}. In the following, we examine these two scenarios in the context of the results presented in Sec. \ref{sec:results} for MAXI J1820+070.

\subsection{Synchrotron emission from a compact jet}
\label{jet-option}
In BH-LMXBs, a coupling between accretion and ejection has been shown to occur \citep{Merloni2003, Falcke2004, Plotkin2013}, with matter ejected primarily through compact, collimated jets launched near the black hole. In the hard X-ray state, a flat radio spectrum typically indicates a jet \citep{Fender01, Fender04, Corbel2004}. This flat or inverted spectrum ($\alpha \sim 0$ to $0.5$, with $F_\nu \propto \nu^\alpha$) often extends to the infrared \citep[e.g.][]{Corbel02,Gandhi11,Rout21,Saikia2019} and results from overlapping self-absorbed synchrotron components at different distances from the black hole.

At higher frequencies (infrared to optical), the jet emits optically thin synchrotron radiation after a spectral break, producing a steeper spectrum ($-1 < \alpha < -0.5$; \citealt{Buxton2004,Gandhi11,Russell2013b}). This appears as an infrared excess above the extrapolated disc emission and is often highly variable \citep{Casella2010,Chaty2011}, with fractional rms amplitudes up to about 20 per cent on timescales of seconds to minutes \citep[e.g., GX 339-4, MAXI J1535-571;][]{Gandhi11,CadolleBel2011,Baglio2018}.

\citet{Baglio2018} showed that the mid-IR variability in MAXI J1535-571 is consistent with the internal shock model \citep{Malzac2013, Malzac2014}, where variability in the accretion flow creates velocity fluctuations in shells of plasma in the jet, which then collide, generating shocks within the jet. These shocks accelerate particles and produce rapidly varying synchrotron emission. Jet internal shock models \citep[][]{Jamil2010,Malzac2013} have also been used to explain the broadband (radio to optical--IR) SEDs and/or fast timing properties (including lags with the X-ray flux) of GX~339--4 \citep{Drappeau2015,Malzac2018}, MAXI~J1836--194 \citep{Peault2019} and GRS~1716-249 \citep{Bassi2020}. As well as strong variations on minute--hour down to sub-second timescales, the jet internal shock model also predicts a $\sim 0.1$ s optical--IR lag with respect to X-ray variability \citep{Malzac2018}, which has been seen in both IR--X-ray and optical--X-ray cross correlations from several BHXBs \citep{Gandhi2008,Gandhi2010,Gandhi2017,Casella2010,Paice2019,Ulgiati2024}.

During the soft X-ray state, jets appear quenched across all wavelengths \citep[e.g.][]{Gallo2003,Homan2005,Baglio2018}. There is evidence that the jet spectral break shifts from IR to radio frequencies during the hard to soft transition, and vice versa during the return to the hard state \citep{Coriat2009,Corbel2013,Russell2013b,Russell.T.2020}, with the jet break frequency correlating with the X-ray hardness \citep{Koljonen2015,Echiburu2024}. The amplitude of the fade/recovery of the IR excess is related to the system inclination, implying that the compact jets are outflowing with Lorentz factors $\sim 1.3$--3.5 \citep{Saikia2019} and at the peak of the hard state, the jet contributes $\sim 90$ per cent and $\sim 50$ per cent of the NIR and optical flux, respectively \citep{Russell2006}.

In MAXI J1820, a prominent IR excess was seen in the hard state \citep[both before and after the soft state;][]{Russell2018ATel11533,OzbeyArabaci2022,Echiburu2024}, and the IR flux was shown to have stronger variations on sub-seconds to minute timescales than the optical flux \citep{Tetarenko2021}. The power spectral density smoothly evolves from optical to radio wavelengths, with a tight correlation between the break in the power spectrum, and the observational frequency \citep{Tetarenko2021}. This variability is consistent with internal shocks in the jet, from optical to radio, and the PDS break is related to the distance downstream in the jet \citep{Tetarenko2021}. The $\sim 0.1$ sec optical lag commonly attributed to the jet was also reported, with the lag increasing slightly with optical wavelength \citep{Paice2019,Paice2021}.

\begin{figure*}
    \centering
    \includegraphics[width=1\linewidth]{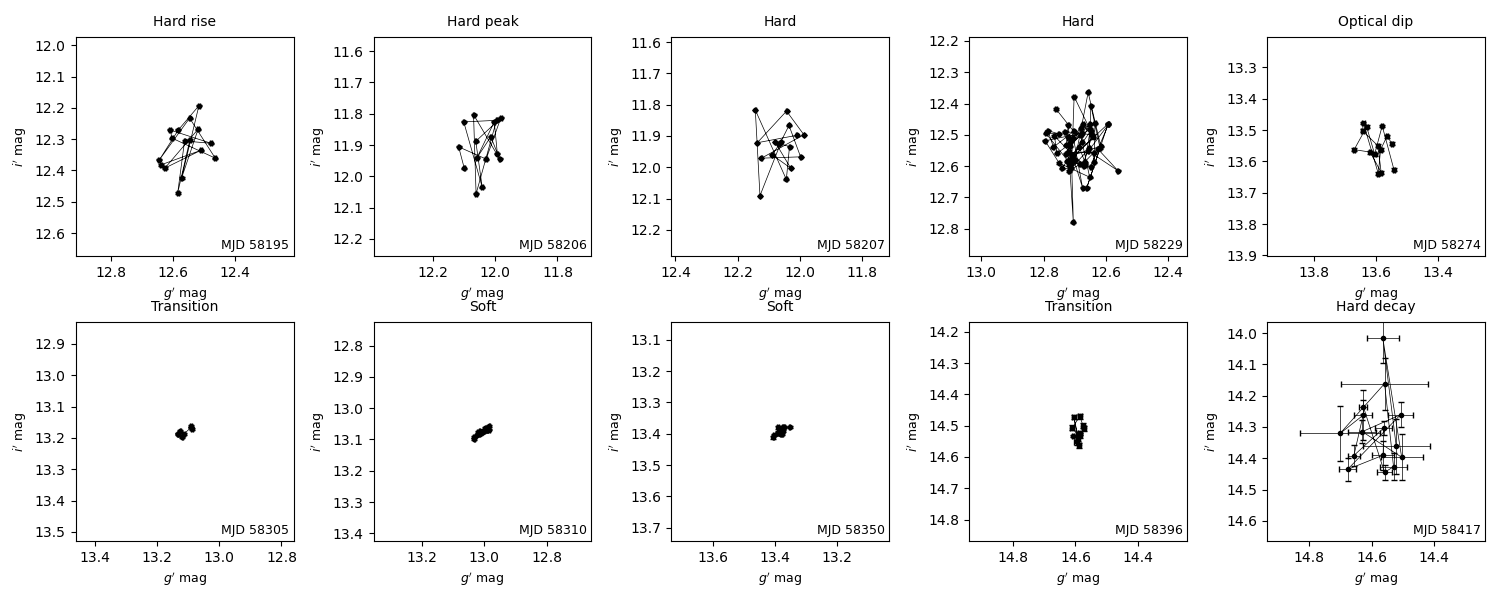}
    \caption{Short-timescale magnitude–magnitude diagrams for MAXI J1820+070 across different phases of its outburst. Each panel shows $g'$ vs. $i'$ magnitudes for a specific MJD (indicated in the lower right corner), with point-to-point connections highlighting the light curve evolution. The different phases of the outburst are indicated on top of each panel, reflecting the spectral state inferred from simultaneous X-ray monitoring.}
    \label{fig:appendix2}
\end{figure*}

Figures \ref{Fig:rms_LCO} and \ref{Fig:rms_AlSadeem} show that optical minute-timescale variability increases with X-ray hardness, likely indicating a stronger contribution from the jet during harder states. In the LCO data, this is particularly clear in the $i'$-band, suggesting that the variable component is more prominent at longer wavelengths. This is also supported by Fig. \ref{fig:appendix2}, where consecutive $g'$ and $i'$-band magnitudes are shown for different outburst stages. Since most observations were composed of consecutive, alternating exposures in $g'$ and $i'$-bands, it was possible to match adjacent magnitudes to see if $g'$ and $i'$ magnitudes are correlated, and to compare their variability amplitudes. In the hard state, the distribution of magnitudes has an oval shape, with stronger variability in the $i'$-band and no clear correlation between the two bands. This pattern is consistent with rapid, flickering-like variability on timescales shorter than the filter-switching cadence of $\sim 2$--3 minutes.

At the hard-to-soft state transition, the variability amplitude decreases and the $g'$ and $i'$ magnitudes begin to rise together. In the soft state, the two bands are well correlated (e.g. on MJD 58310 and 58350), although the fractional rms is low (typically below 1--2 per cent), and the variations seem to be dominated by long-term trends, rather than short term flickering (Fig.~\ref{fig:appendix1}). This is in line with expectations for a disc-dominated emission scenario.

Figure~\ref{Fig:xray_rms} shows a positive correlation between the optical and X-ray fractional rms, which is stronger for shorter X-ray binning (5s). The correlation weakens significantly at longer binning (60s), where the X-ray rms flattens below 10 per cent. This suggests that the strongest X-ray variability occurs on short timescales. The link between short-timescale variability in the optical and in the emission of the corona, both of which increase with spectral hardness, supports a linked origin. According to the internal shock model, variability in the emission of the corona can introduce velocity fluctuations that generate internal shocks in the jet. These would then produce fast synchrotron variability in the optical, particularly in the $i'$-band where the jet is expected to contribute more strongly, due its red (optically thin) spectrum, and the bluer disc emission becoming progressively stronger at shorter wavelengths.

In the soft state, the suppression of the jet leads to minimal minute-timescale variability, as observed. In addition to the lower amplitude, the variability becomes slower and better correlated between $g'$ and $i'$-bands.
This is consistent with low-level disc fluctuations or reprocessing of X-ray radiation in the extended outer disc \citep{Shidatsu2019}.

\begin{center}
\begin{figure}
\includegraphics[scale=0.46]{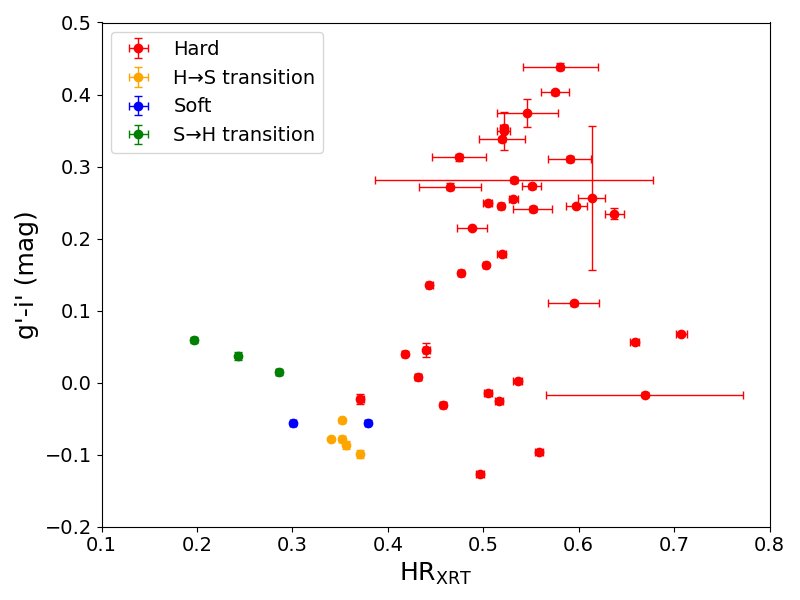}
\caption{ $\rm HR_{XRT}$ vs. $g'$-$i'$ color from the LCO monitoring. Different X-ray spectral states are indicated with different colours following the classification of \citealt{Shidatsu2019, Fabian2020}. }
\label{fig:col}
\end{figure}
\end{center}

To explore this further, we examined the $g'-i'$ colour as a function of X-ray hardness (Fig. \ref{fig:col}). The system remains consistently blue ($g'-i' \lesssim 0.1$) during the soft state and state transitions \citep{Shidatsu2019, Fabian2020}, in line with expectations for a disc-dominated emission scenario. During harder epochs ($HR_{\rm XRT} \gtrsim 0.4$), the colour becomes more variable, with both blue and red values observed. Redder colours tend to appear at low optical fluxes, which also correspond to the hard state. While there is no global trend, the soft and hard states display clearly distinct colour behaviour. In the hard state, the variability is dominated by fluctuations on shorter timescales than the filter switching (Fig. \ref{fig:appendix2}), resulting in uncorrelated $g'$ and $i'$ magnitudes, and hence, highly variable $g'-i'$ colours.

The optical fractional rms variability is greater than a few per cent in all hard state observations (including the re-brightening in 2019; Fig.~\ref{lc_fig} and below 1--2 per cent in the soft state. The highest rms (5--10 per cent) was recorded during the first $\sim 40$ days in the initial hard state, and during the decaying hard state after the soft state. Interestingly, these were epochs when the jet spectral break was at the highest frequencies ($> 5\times 10^{12}$ Hz), with the most prominent IR-excess \citep{Echiburu2024}.
The persistent blue colour during the soft and transition states is consistent with the quenching of the jet, as reported by \citet{Echiburu2024}. The hard to soft transition, which occurred between MJD 58303.5 and 58310.7 \citep{Yang2025} and during which the radio jet quenched \citep{Bright2020}, also coincides with a drop in optical $g'-i'$ colour and variability in our data (orange points in Fig. \ref{fig:col}), reinforcing the idea that suppression of the jet leads to reduced optical variability.


These results suggest that the observed minute-timescale optical variability originates from the jet in the hard state. When the jet is quenched, such variability disappears, and the low residual variability may reflect slow disc fluctuations. This interpretation is further supported by the optical–X-ray fractional rms correlation, particularly in the $i'$-band. The positive correlation implies that the variable optical component is stronger at longer wavelengths, consistent with jet-dominated synchrotron emission.

This scenario also provides a natural explanation for the outlier observed on MJD 58305 in the rms correlation plot (lower right points in Fig.~\ref{Fig:xray_rms}). This date corresponds to the hard-to-soft transition, when the jet is believed to have been quenched \citep{Bright2020}. Although the X-ray fractional rms remains elevated, indicating ongoing variability in the emission of the corona, the optical rms drops sharply. If the jet dominates fast optical variability in the hard state, its suppression would lead to a breakdown in this correlation.

Overall, our results support a picture in which the jet plays a dominant role in producing fast optical variability during the hard state, while the soft state is characterized by stable disc emission, in line with expectations for a disc-dominated emission scenario. This is consistent with previous optical/X-ray studies of this source, whereby the jet was claimed to make a contribution to the optical flux \citep{shidatsu2018,Shidatsu2019,Yang2025,Bright2025} and variability \citep{Paice2019,Paice2021,Tetarenko2021} in the hard state. The optical fractional rms correlates well with the jet emission in the IR. The smoothly evolving power spectral shape from optical to radio \citep[fig. 5 in][]{Tetarenko2021}, where the break in the PDS correlates very well with frequency, was measured on MJD 58220, when our rms was $\sim 5$--6 per cent \citep[$\sim 20$ per cent rms was measured at optical wavelengths, integrated over all timing frequencies, by][]{Tetarenko2021}. This shows that the majority of the variability (over all timescales) is produced by the jet, at least on that particular date in the hard state.

\subsection{The hot magnetised inflow}

To account for the complex phenomenology observed in BH-LMXBs, including rapid variability across OIR to X-ray wavelengths, inter-band correlations, and broadband spectral behavior, a modification to the standard thin accretion disc model has been proposed. This framework introduces a radiatively inefficient, magnetized, synchrotron-emitting hot flow that occupies the innermost regions of the accretion geometry, extending inward toward the black hole, potentially down to the innermost stable circular orbit (ISCO; \citealt{Veledina2013}). The spectral energy distribution of this hot accretion flow, and any advection dominated accretion flow (ADAF) or radiatively inefficient accretion flow (RIAF) in the OIR ranges depends sensitively on the radial profiles of key physical quantities, such as the magnetic field strength and electron density \citep{Esin1997,BlandfordBegelman1999}. The inclusion of non-thermal electrons, as incorporated in hybrid numerical models of hot flows \citep[e.g.,][]{PoutanenVeledina2014}, further modifies the emitted spectrum, typically hardening the OIR spectral energy distribution (SED) slope and extending emission to higher energies.

In this scenario, the lower-energy cut-off of the synchrotron spectrum marks the outer boundary of the hot flow (often on the order of several hundred gravitational radii) and typically falls within the OIR domain. At higher energies, in the ultraviolet and optical bands, the spectrum transitions from partially self-absorbed synchrotron to optically thin synchrotron, and eventually to a Comptonized component. It has been suggested that this hot flow could make a substantial contribution from the IR to X-ray flux, particularly during the hard state \citep{Veledina2013}, although the standard accretion disc is still expected to dominate at optical to UV wavelengths.

A feature of the hot flow model is its prediction for short-timescale variability: optical/NIR photons generated by the synchrotron process in the hot inner flow are expected to be anti-correlated with X-ray fluctuations, due to a spectrum that pivots in the UV \citep{Veledina2011}. Here, X-rays arise via Compton upscattering by the same population of hot electrons that produce the optical emission. An increasing mass accretion rate results in brighter X-rays, and reduced optical synchrotron emission due to the increased synchrotron self-absorption. This leads to a characteristic anti-correlation and time delay, with X-ray flares lagging optical dips, and was developed to account for the `precognition dip' in the optical/X-ray cross-correlation function \citep{Veledina2011}, a feature that had already been explained by a jet `energy reservoir' model \citep{Malzac2004}. Such lag signatures contrast with those expected from standard reprocessing, in which optical--IR fluctuations would be produced by a smeared, lagged response to X-rays variations \citep[][]{OBrien2002,hynes2005}.

While the hot flow could produce detectable synchrotron emission at optical wavelengths, its spectrum is expected to be flat, $\alpha \sim 0 \pm 0.5$, and become fainter at longer wavelengths into the IR \citep{Veledina2013}. The observed optical--IR flux spectrum in the hard state is red \citep[$\alpha \ll 0$;][]{Echiburu2024}, with a peak of the flux density in the mid-IR.  The mid-IR excess and red SED would imply a very large inner disc truncation radius and challenge the hot flow scenario. The optical rms variability is highest when the mid-IR is most prominent (section \ref{jet-option}). The variability properties, in particular the fractional rms and the frequency of the break in the PDS, correlates extremely well with wavelength, from optical to radio \citep{Tetarenko2021}. Since the above observational constraints are inconsistent with expectations for the hot flow, it seems unlikely that this component is responsible for the optical rms variability reported in this work.

\cite{Paice2019,Paice2021} studied the evolution of the optical/X-ray cross-correlation functions (CCFs) using simultaneous data of MAXI J1820 in the initial hard state (MJD 58193--58276). They found that during this time, a sharp $\sim 0.1$ sec lag and a precognition dip were superimposed in the CCFs, and red flares were observed. The $\sim 0.1$--0.2 sec lag, which has been established as a feature of optical--IR jet emission \citep{Gandhi2017}, was strong near the peak of the outburst and weakened later in the hard state. The lag increased at longer optical wavelengths, which is consistent with those longer wavelengths originating from distances further out in the jet, and the progression of matter producing shorter then longer wavelength emission \citep{Paice2021}. The precognition dip, which could be caused by the hot flow, became more prominent during later stages in the hard state, and became broader and shifted to positive lags (optical dips lagging X-ray flares). A model was able to reproduce the CCFs, QPOs and phase lags if the jet dominates the variability at the lowest ($<0.1$ Hz) and the highest ($>1$ Hz) frequencies, with the hot flow contributing in between \citep{Paice2021}. This is consistent with the findings of \cite{Tetarenko2021}, which found the integrated variability (over short and long timescales) increases at longer wavelengths and so has a jet origin, and with the rms variability reported in this work, which is red, and probes relatively long ($\sim$ minute) timescales ($<0.01$ Hz). In the \cite{Paice2021} model, the hot flow made a relatively stronger contribution to the (anti-correlated) variability after MJD $\sim 58225$, which is roughly when our rms variability drops below $\sim 5$ per cent. This residual, low-level optical variability could therefore be attributed to cyclo-synchrotron emission from the hot flow.

\subsection{Low-flux variability}

\begin{figure}
    \centering
    \includegraphics[width=1\linewidth]{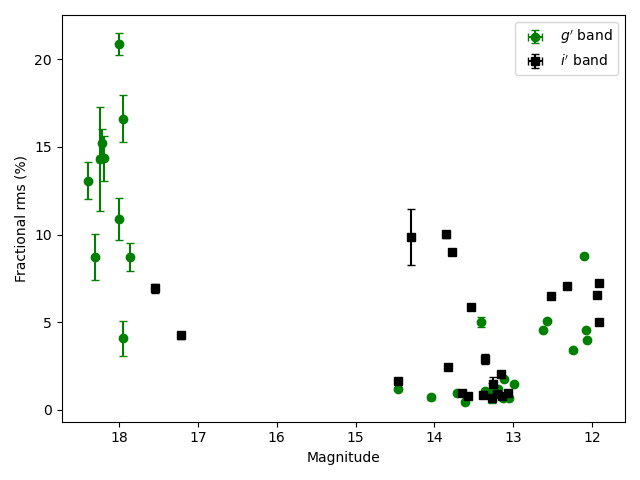}
    \caption{Fractional rms variability as a function of magnitude in the LCO $g'$ and $i'$ bands (green and black points, respectively). }
    \label{fig:quiescent_varuiability}
\end{figure}

In Fig. \ref{fig:quiescent_varuiability}, we show the fractional rms variability as a function of magnitude in the $g'$ and $i'$ bands. The magnitudes have been averaged over closely spaced epochs and matched to the rms measurements at the nearest MJDs. As the source faded during outburst decay, the fractional rms variability decreases with increasing flux (i.e., decreasing magnitude). However, we observe a striking increase in the optical variability at very low fluxes, corresponding to the (nearly) quiescent state.

In particular, the $g'$-band rms reaches values above $20\%$ at magnitudes $\gtrsim 18$, indicating significant variability in the optical even when the source is less active. This level of quiescent optical variability is reminiscent of the behavior reported for the BH LMXB \textit{Swift} J1357.2–0933, which exhibited large-amplitude, short-timescale flares in quiescence \citep{Shahbaz2013}. In both cases, the high rms suggests that the emission mechanism in quiescence is dominated by a highly variable component, possibly synchrotron radiation from a weak jet or magnetic reconnection events in the accretion flow.

\section{Optical spectroscopy}\label{spec}
MAXI J1820+070 was observed spectroscopically with the Cassini 1.5m telescope at the Loiano Astronomical Observatory (Italy) using the Bologna Faint Object Spectrograph \& Camera (BFOSC\footnote{\url{https://www.oas.inaf.it/wp-content/uploads/2019/05/BFOSC\_en\_UM\_2001.pdf}}) instrument on 2018 June 8 (MJD 58277), July 17 (MJD 58316), July 18 (MJD 58317), November 9 (MJD 58431), as well as on 2019 September 
20 (MJD 58746). Observations were also acquired with the 2.1m telescope of the Observatorio Astrofísico Guillermo Haro (OAGH) in Cananea (Mexico), using a Boller \& Chivens spectrograph, on 2018 April 15 (MJD 58223), and with the 2.1m telescope at the Observatorio Astronómico Nacional San Pedro Mártir (OAN-SPM; Mexico), also equipped with a Boller \& Chivens spectrograph, on 2018 August 8 (MJD 58338). Exposure times ranged from 180 to 1800 s. A log of the observations is reported in Table \ref{tab:log_spec}

The BFOSC spectra were acquired with the Grism \#4 and a slit 
width of 2$''$, providing a nominal spectral coverage of the
3500-8500 \AA\, range and a resolution of $\sim$10 \AA; the 
OAGH spectra were secured with a 150 lines/mm grating and a slit width of 
300 $\mu$m ($\sim$2$\farcs$5), affording a resolution of $\sim$15 \AA\, 
in the 3600-7300 \AA\, range, whereas the OAN-SPM ones were secured with a 
300 lines/mm grating, a slit width of 2$\farcs$5 and a resolution of 
$\sim$8 \AA, covering the wavelength range between 3700 and 8400 \AA.

Data reduction was carried out using standard procedures for both bias subtraction and flat-field correction with IRAF\footnote{IRAF is the 
Image Analysis and Reduction Facility made available to the astronomical 
community by the National Optical Astronomy Observatories, which are operated by AURA, Inc., under contract with the U.S. National Science Foundation. It is available at {\tt http://iraf.noao.edu/}}, and the spectra were optimally 
extracted according to the procedure of \citet{Horne1986}.
The wavelength calibration was carried out with helium-argon (Cassini and
OAGH data) or neon-helium-copper-argon (OAN-SPM) lamps. The flux calibration was performed using catalogued spectroscopic standard stars within IRAF.

\begin{center}
\begin{figure}
\includegraphics[scale=0.4]{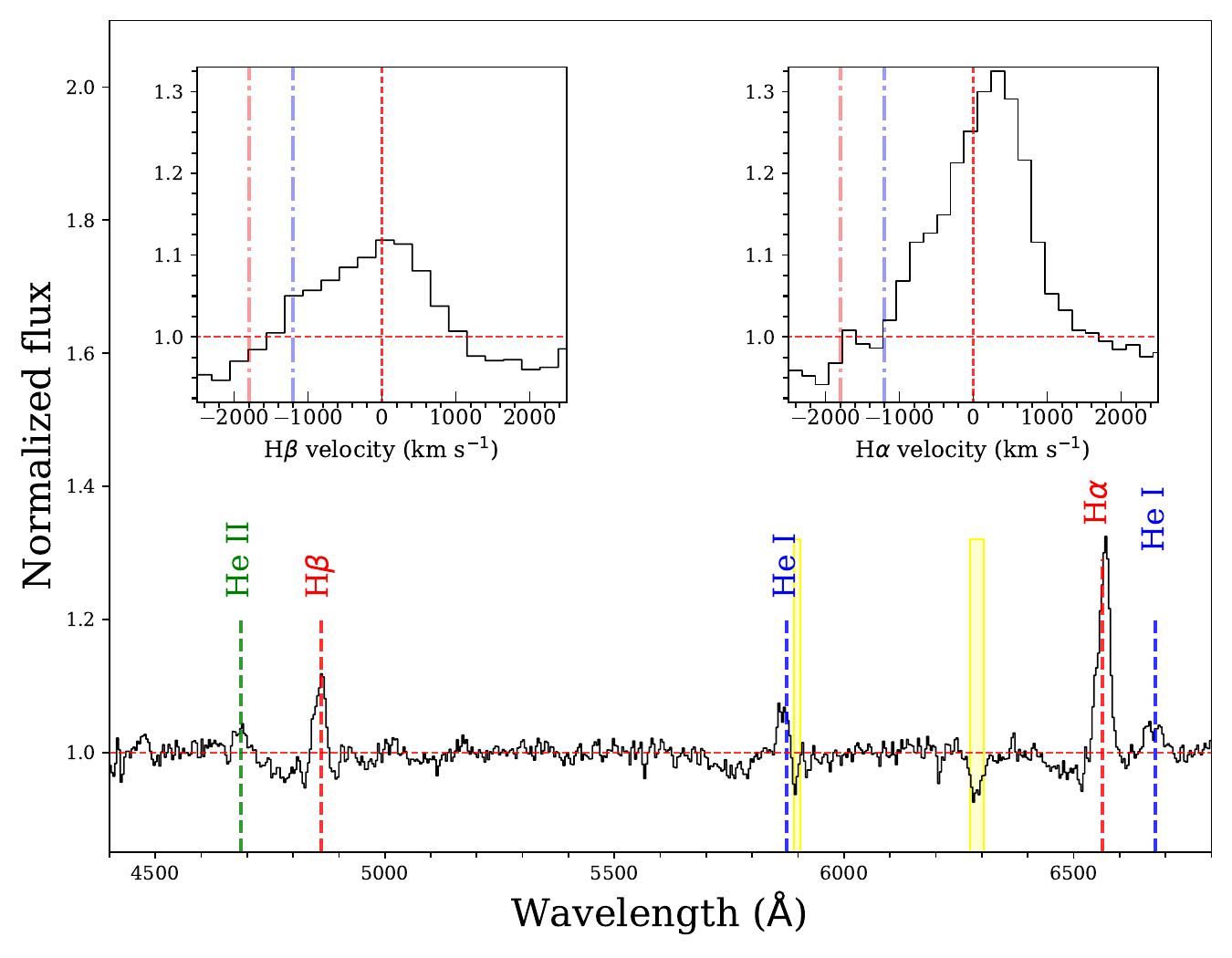}
\caption{Full normalized spectrum acquired on 2019 September 20, during the hard X-ray spectral state. The position of some of the most relevant emission lines is highlighted with dashed blue (He I 5876\AA\, and 6678\AA), red ($H_{\alpha}$, $H_{\beta}$) and green (He II) lines. In yellow, the Na interstellar doublet is marked. The two insets show the zoom of the normalized spectrum in velocity scale, for the H$_{\beta}$ and H$_{\alpha}$ lines. In the insets, we marked with dotted-dashed lines the wind velocities (blue-edge) measured in the system in previous studies.}
\label{spectrum_full}
\end{figure}
\end{center}

In line with previous spectroscopic works on this source \citep[][]{Tucker2018, MunozDarias2019, Sanchez2020}, all of our spectra exhibit numerous emission lines, featuring the Balmer series along with multiple He I and He II transitions (Fig. \ref{fig:spectra}). Following a visual inspection of all the observing epochs, we will focus our analysis on the best spectrum taken during the hard state, during the second re-brightening, which suggests the presence of both disc and wind components.

\subsection{Wind signatures during the second re-brightening}

Fig. \ref{spectrum_full} shows in detail the average spectrum taken during the second re-brightening, on 2019 September 20 (MJD 58746). This is the best quality spectrum that we obtained while the source was in the hard state, due to the large exposure time (the spectrum is the average of three 1800s spectra; Tab. \ref{tab:log_spec}). This is the only epoch in which both the $H_{\alpha}$ and $H_{\beta}$ lines are distorted (i.e. they show a redshifted core; see inset in Fig.~\ref{spectrum_full}), as was observed in the hard state of the main outburst \citep[e.g.][]{MunozDarias2019}, sometimes simultaneously with P-Cyg profiles in other transitions. In the same way as P-Cyg line profiles, these distorted (red-shifted) profiles can result from partial absorption of the blue part of the emission component, but in this case the absorption is not strong enough to significantly dip below the continuum level. Our soft-state spectra do not show such signatures, in agreement with \citet{MunozDarias2019}, and instead exhibit symmetric, double-peaked profiles. Furthermore, we notice the presence of a possible absorption trough in the blue wing of the H$\alpha$ line (see inset in Fig.\ref{spectrum_full}). The velocity associated with this trough is unconstrained, especially considering that broad absorption components are also present (see Sec. \ref{additional_features} below). However, it seems broadly consistent with the wind velocity measured in the past ($\sim 1200$–$1800$ km/s; blue and red dashed-dotted lines in the inset of Fig.~\ref{spectrum_full}; \citealt{MunozDarias2019, Sanchez2020}).

Cold winds in MAXI J1820+070 were first clearly identified during the hard state of the 2018 main outburst through high-resolution spectroscopy by \citet{MunozDarias2019}. These features were absent in the soft state, though a near-IR wind was later reported by \citet{Sanchez2020}. Similar low-ionisation winds have been observed in several other BH-LMXBs, including V404 Cyg \citep{MunozDarias2016}, V4641 Sgr \citep{MunozDarias2018}, GRS 1716–249 \citep{cuneo20}, MAXI J1803–298 \citep{matasanchez2022}, and MAXI J1348–630 \citep{Panizo2022}, suggesting that they are a common feature of black hole accretion. Perhaps as a result of our limited spectral resolution, definitive signatures of optical--NIR winds (e.g. P-Cygni profiles) are not clearly resolved in our data. However, the distorted profiles observed in H$\alpha$ and H$\beta$ suggest that cold winds were likely present during this rebrightening (see also \citealt{Sanchez2020}), strengthening the case for the presence of these outflows during low-luminosity hard states. This further supports the scenario in which jets and winds can coexist in X-ray binaries.

On the other hand, the absence of obvious wind signatures in our soft-state spectra may instead reflect higher ionisation in the accretion disc, which would suppress low-ionisation features in this phase \citep[see also][]{MunozDarias2019}. To investigate this possibility, we examined the ratio between the equivalent widths (EW) of the He II 4686 \AA\ and $H_{\alpha}$ emission lines as a function of the hardness ratio HR$_{\rm BAT/MAXI}$. This ratio is systematically higher during soft states compared to hard states, consistent with enhanced disc ionisation. However, systematic effects may contribute, and repeating the analysis with $H_{\beta}$ in place of $H_{\alpha}$ did not reveal a similar correlation, likely due to the larger uncertainties affecting the $H_{\beta}$ EW measurements.

\subsection{Additional spectral features during the reflare}
\label{additional_features}
In addition to the aforementioned distorted emission profiles, broad absorptions components are found underlying the emission lines in the spectrum (H$\alpha$, H$\beta$, He~I). Such features are not uncommon and have been observed, typically at low luminosity, in other systems. While their origin is unclear, they might be related to the disc becoming optically thick and behaving like a stellar atmosphere (see e.g. \citealt{Jimenez-Ibarra2019, Dubus2001}, and references therein). Finally, He I lines are double-peaked, suggesting for their origin in the accretion disc. The He I 5876\AA\, line looks blue-shifted, but this effect could be due to the presence of the Na interstellar doublet rather than to the presence of an outflow. He I 6678\AA\, is symmetric, broad and double-peaked, as expected for an accretion disc origin.

\section{Conclusions}

In this work, we present the results of an optical campaign performed on the BH-LMXB MAXI J1820+070 during its main 2018 outburst and the following re-brightenings. The campaign involves a systematic photometric monitoring with the telescopes of the Las Cumbres Observatory Network and the telescope of the Al Sadeem Observatory in the UAE, together with low-resolution spectroscopy performed across different spectral states in the same time window using the 2.1 m telescope at the Observatorio Astronómico Nacional San Pedro Mártir,
with the 2.1 m telescope of the Observatorio Astrofísico Guillermo Haro in Cananea (M\'{e}xico) and with the 1.5-m G.D. Cassini telescope of the Loiano Astronomical Observatory (Italy). Within the photometric monitoring, we specifically focus on time-resolved light curves obtained intermittently throughout the source active period, with a timescale of approximately $\sim$1 minute. The main results we obtained are summarized below.

\begin{itemize}
    \item The minute-timescale optical light curves display substantial variability depending on the X-ray spectral state. By evaluating the hardness ratio using the $Swift$/BAT count rate for the hard band and the MAXI count rate for the soft band, we find a strong positive correlation between the hardness ratio and the optical fractional rms, with the latter increasing alongside hardness, especially at lower optical frequencies. Similarly, we observe a strong positive correlation between the optical fractional rms and the hardness ratio (calculated as the 2--10 keV to 0.5--2 keV $Swift$/XRT count rate) but only when the hardness ratio exceeds $\sim0.35$. These findings suggest that variable synchrotron emission from the compact jet likely drives the observed strong minute-timescale variability. According to accretion-ejection coupling, the jet contributes variable emission during hard states, while it is quenched in soft states.    
    \item We show that there is a strong correlation between optical fractional rms and the 0.01--64 Hz $Swift$/XRT fractional rms amplitude evaluated in the 2--10 keV energy range. This correlation is particularly strong for the optical $i'$-band. This result is in agreement with the jet scenario, and in particular with the internal shock model \citep{Malzac2013, Malzac2014}, according to which a variable accretion flow is responsible for injecting velocity fluctuation at the base of the jet, therefore producing shocks that generate variable synchrotron emission in the jet, observable in infrared and optical light curves at different timescales (from sub-seconds to minutes). In the dimmest hard states, the residual very low-level optical variability could instead be attributed to cyclo-synchrotron emission from the hot flow, consistent with the results of \citet{Paice2021}.
    \item Our low-resolution optical spectroscopy did not reveal clear signs of winds, such as line asymmetries or P-Cygni profiles. Some hard-state spectra show hints of P-Cygni profiles, though none appear in the soft-state spectra. This aligns with previous findings on the source and further supports the presence of optical winds during hard states in MAXI~J1820+070. The absence of optical wind signatures in soft states could be related to over-ionisation of the outflow, as suggested by the positive correlation between the ratio of the equivalent widths of the He\textsc{ii}~4686\AA\ and H$\alpha$ emission lines and the X-ray hardness.

\end{itemize}

\begin{acknowledgements}
We thank the referee, for valuable comments which helped improving our manuscript.
We thank Daniel Bramich for valuable discussions on variability estimates and for developing the XB-NEWS pipeline, on whose results this work is based. 
We moreover thank Silvia Galleti, Ivan Bruni, Antonio De Blasi and Roberto Gualandi of the Loiano Observatory for the assistance at the telescope. 
MCB is supported by the INAF-Astrofit fellowship.
 FMV is supported by the European Union’s Horizon Europe research and innovation programme with the Marie Sk\l{}odowska-Curie grant agreement No. 101149685. 
KA, DMR, SR and PS are supported by Tamkeen under the NYU Abu Dhabi Research Institute grant CASS.
NM acknowledges financial support through ASI-INAF 2017-14-H.0 agreement (PI: T. Belloni). T.M.-D. acknowledges support by the Spanish \textit{Agencia estatal de investigaci\'on} via PID2021-124879NB-I00. 
This publication is based on data collected at the Observatorio Astroíısico Guillermo Haro (OAGH), Cananea, Sonora, México, operated by the Instituto Nacional de Astrofísica, Óptica y Electrónica (INAOE). This paper is based upon observations carried out at the Observatorio Astronómico Nacional on the Sierra San Pedro Mártir (OAN-SPM), Baja California, México.
\end{acknowledgements}

\bibliography{bibliography}{}
\bibliographystyle{aasjournal}

\newpage
\begin{appendix}
    
\section{LCO short-timescale light curves}\label{lco_appendix}
\begin{figure}[!h]
    \centering
    \includegraphics[width=1\linewidth]{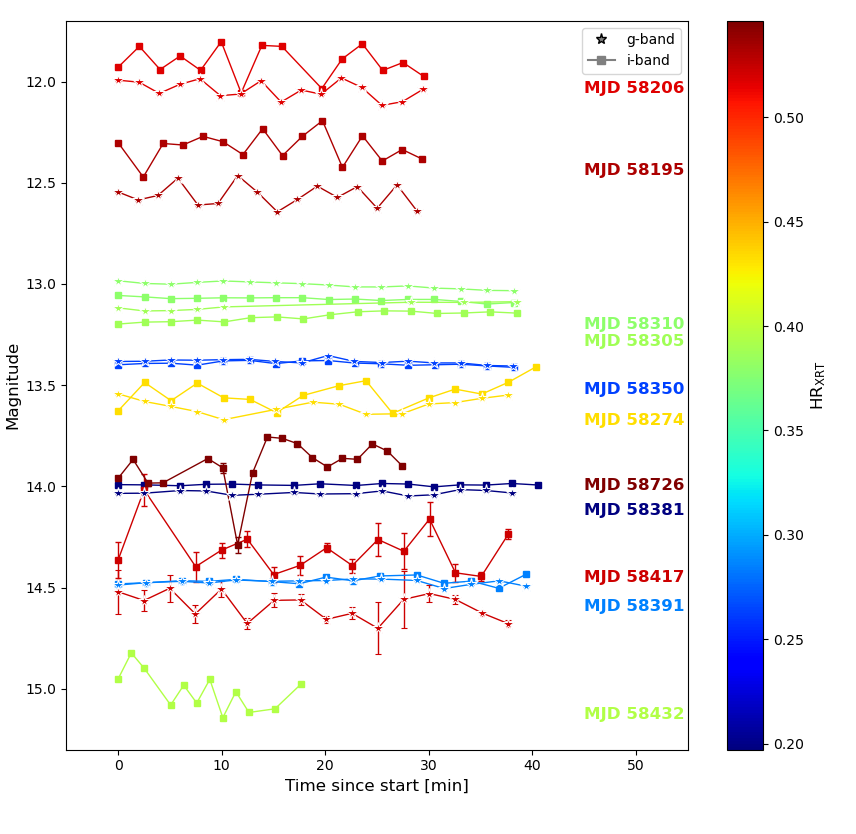}
    \caption{Selection of short-timescale light curves of MAXI J1820+070 from the campaign presented in this work. The X-ray spectral hardness is color-coded, as indicated by the color bar (red = hard; blue = soft). The $g'$- and $i'$-band light curves are shown as stars and squares, respectively. Time is measured in minutes since the start of each light curve.}
    \label{fig:appendix1}
\end{figure}

A selection of short-timescale light curves of MAXI J1820 throughout the 2018-2019 main outburst and reflares are shown in Fig. \ref{fig:appendix1}. Details of the light curves (MJDs, exposure times and telescope configuration) can be found in Tab. \ref{log_LCO}.

\begin{table*}
\caption{Log of the fast-timing $i'$ and $g'$-band LCO observations published in this work. In the Site columns, 1m- and 2m- A, C, S, H, T indicate the 1m and 2m telescopes in the LCO nodes located in Australia, Chile, South Africa, Hawaii and Texas, respectively.}     
\label{log_LCO}      
\centering                       
\begin{tabular}{c c c c c | c c c c c}       
\hline               
Epoch (UT) & MJD & Site & Filter & Exposure & Epoch (UT) & MJD & Site & Filter & Exposure \\   
\hline  
2018-03-18  & 58195.78472 & 2m-A & $g'$ & 16x20s &  2018-08-20  & 58350.13563  & 1m-C & $g'$ & 16x20s \\
            & 58195.78399 & 2m-A & $i'$ & 16x20s &              & 58350.13474 & 1m-C & $i'$ & 16x20s \\
2018-03-26  & 58203.77896 & 2m-A & $g'$ & 5x20s  &  2018-08-31  & 58361.10082 & 1m-C & $g'$ & 16x20s \\
            & 58203.77575 & 2m-A & $i'$ & 8x20s  &              & 58361.09861 & 1m-C & $i'$ & 16x20s \\
2018-03-29  & 58206.74406 & 2m-A & $g'$ & 16x20s &  2018-09-11  & 58372.04031 & 1m-C & $g'$ & 14x20s\\
            & 58206.72588 & 2m-A & $i'$ & 16x20s &              & 58372.03910 & 1m-C & $i'$ & 16x20s \\
2018-03-30  & 58207.54593 & 2m-H & $g'$ & 16x20s &  2018-09-20  & 58381.77835 & 1m-S & $g'$ & 15x20s \\
            & 58207.54522 & 2m-H & $i'$ & 16x20s &              & 58381.77935 & 1m-S & $i'$ & 16x20s \\
2018-04-21  & 58229.28690 & 2m-H & $g'$ & 16x20s &  2018-09-30  & 58391.39042 & 1m-S & $g'$ & 16x20s \\
            & 58229.39152 & 1m-C & $g'$ & 48x20s &              & 58391.38950 & 1m-S & $i'$ & 16x20s \\
            & 58229.28601 & 2m-H & $i'$ & 16x20s &  2018-10-14  & 58405.09079 & 1m-T & $i'$ & 20x30s \\
            & 58229.38908  & 1m-C & $i'$ & 48x20s &  2018-10-26  & 58417.09342 & 1m-T & $g'$ & 16x20s \\
2018-06-05  & 58274.89816 & 1m-S & $g'$ & 14x20s &              & 58417.09318 & 1m-T & $i'$ & 16x20s \\
            & 58274.89783 & 1m-S & $i'$ & 16x20s &  2018-11-10  & 58432.20869 & 2m-H & $i'$ & 12x60s \\
2018-06-26  & 58295.49099 & 1m-A & $g'$ & 16x20s &  2019-02-26 & 58540.77288 & 2m-A & $i'$ & 20x60s \\
            & 58295.49009 & 1m-A & $i'$ & 16x20s &  2019-05-08 & 58611.66464  & 2m-A & $i'$ & 20x60s \\
2018-07-06  & 58305.64781 & 1m-A & $g'$ & 16x20s &  2019-05-18 & 58621.49958 & 2m-H & $g'$ & 6x20s \\
            & 58305.64711 & 1m-A & $i'$ & 16x20s &             & 58621.60684  & 2m-A & $g'$ & 6x20s \\
2018-07-08  & 58307.94670 & 1m-S & $g'$ & 7x20s  &  2019-05-19 & 58622.410632 & 2m-H & $g'$ & 6x20s \\ 
            & 58307.94818  & 1m-S & $i'$ & 6x20s  &  2019-05-20 & 58623.69558 & 2m-A & $g'$ & 6x20s \\
2018-07-10  & 58309.64963 & 1m-A & $g'$ & 16x20s &  2019-05-22 & 58625.50689 & 2m-H & $g'$ & 6x20s \\
            & 58309.64874 & 1m-A & $i'$ & 16x20s &             & 58625.60771 & 2m-A & $g'$ & 6x20s \\
2018-07-11  & 58310.64080 & 1m-A & $g'$ & 16x20s &  2019-05-23 & 58626.61230 & 2m-A & $g'$ & 6x20s \\
            & 58310.63991 & 1m-A & $i'$ & 16x20s &  2019-05-24 & 58627.480131 & 2m-H & $g'$ & 6x20s \\
2018-07-20  & 58319.97691 & 1m-S & $g'$ & 16x20s &             & 58627.60675 & 2m-A & $g'$ & 6x20s \\
            & 58319.97602 & 1m-S & $i'$ & 16x20s &  2019-05-25 & 58628.41366 & 2m-H & $g'$ & 6x20s \\
2018-07-31  & 58330.20467 & 1m-C & $g'$ & 16x20s &  2019-08-31 & 58726.15288 & 1m-T & $i'$ & 20x60s \\
            & 58330.20379 & 1m-C & $i'$ & 16x20s & & & & & \\
2018-08-13  & 58343.13963 & 1m-C & $g'$ & 16x20s & & & & & \\ 
            & 58343.13786 & 1m-C & $i'$ & 16x20s & & & & & \\ 
\hline
\hline                             
\end{tabular}
\end{table*}

\section{Al Sadeem Observatory short-timescale light curves}

\begin{center}
\begin{figure}[!h]
\includegraphics[scale=0.415]{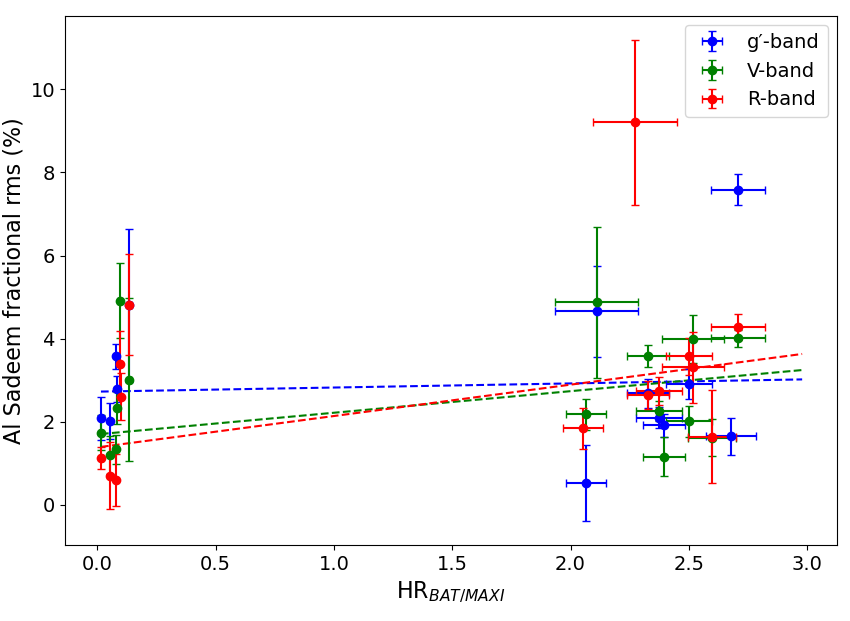}
\includegraphics[scale=0.4]{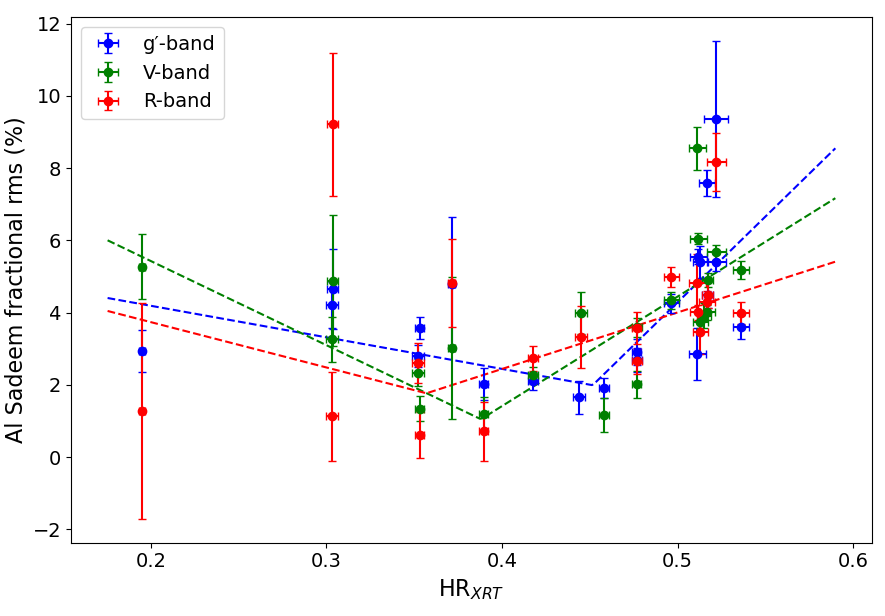}
\caption{\textit{Top}: Fractional rms vs. Hardness ratio calculated as the BAT flux/MAXI flux and the optical fluxes obtained at the Al Sadeem Observatory in the $g'$, $V$ and $R$ bands (blue, green and red points, respectively). Linear fits are represented with color-coded dashed lines. Results of the fits are reported in Tab. \ref{tab:correlations}; \textit{Bottom}: Fractional rms vs. Hardness ratio calculated as the 2-10 keV \textit{Swift}/XRT flux / the 0.5-2 keV \textit{Swift}/XRT flux and the optical fluxes obtained at the Al Sadeem Observatory in the $g'$, $V$ and $R$ bands. Results of the fits are reported in Tab. \ref{tab:correlations_XRT}.}
\label{Fig:rms_AlSadeem}
\end{figure}
\end{center}

A complete log of the short-timescale light curves obtained at the Al Sadeem Observatory can be found in Tab. \ref{log_AlSadeem}.
In Fig. \ref{Fig:rms_AlSadeem}, the correlation between the fractional rms of the optical light curves and the HR ($HR_{\rm BAT/MAXI}$ and $HR_{\rm XRT}$) is shown.

\begin{table*}
\caption{Log of the fast-timing blue-, green- and red-band Al Sadeem Observatory observations published in this work.}            
\label{log_AlSadeem}      
\centering                       
\begin{tabular}{c c c c  | c c c c}       
\hline               
Epoch (UT) & MJD  & Filter & Exposure & Epoch (UT) & MJD &  Filter & Exposure \\   
\hline  
2018-03-29  & 58206.98997     & Blue & 49x30s    &   2018-07-02  & 58301.77580 & Blue & 23x60s   \\
            & 58206.96095     &  Green  & 50x30s &   	       & 58301.75649 & Green & 20x60s  \\
            & 58206.92394     &  Red   & 50x30s  &   	       & 58301.73715 & Red & 21x60s    \\
2018-03-30  & 58207.99351  & Blue & 50x30s &   2018-07-07  & 58306.97089 & Blue & 22x60s   \\
            & 58207.95117   & Green &50x30s &   	       & 58306.94795 & Green & 22x60s  \\
            & 58207.92437   & Red & 50x30s  &   	       & 58306.92155 & Red & 24x60s   \\
2018-03-31  & 58208.92157   & Red & 50x30s  &   2018-07-10  & 58309.77824 & Blue & 25x60s   \\
2018-04-01  & 58209.03725     & Blue & 50x30s &   	       & 58309.77824 & Green & 24x60s  \\
            & 58209.01249   & Green & 50x30s&   	       & 58309.72710 & Red &  25x60s   \\
            & 58209.96373   & Blue & 15x30s &   2018-07-13  & 58312.90145  & Blue & 2x60s5   \\
            & 58209.95550   & Green & 15x30s&   	       & 58312.87719 & Green & 23x60s  \\
2018-04-04  & 58212.98560    & Blue & 50x30s &   	       & 58312.85446 & Red & 25x60s    \\
            & 58212.93575    & Red & 50x30s  &   2018-07-15  & 58314.97037  & Blue & 24x60s   \\
2018-04-05  & 58213.01037  & Green & 50x30s &   	       & 58314.94488 & Green & 23x60s  \\
            & 58213.98375  & Blue & 50x30s &   	       & 58314.92290 & Red & 24x60s    \\
            & 58213.95643   & Green &40x30s &   2018-07-18  & 58317.90586 & Blue &  25x60s  \\
2018-04-06  & 58214.94162     & Blue & 20x60s &   	       & 58317.88311 & Green & 25x60s  \\
            & 58214.97972   & Green &20x60s &   	       & 58317.85960 & Red & 25x60s    \\
            & 58214.96120    & Red & 20x60s  &   2018-07-21  & 58320.89951 & Green & 22x60s  \\
2018-04-07  & 58215.93577   & Blue & 20x30s &   	       & 58320.87391 & Red & 25x60s    \\
            & 58215.92540     & Green &20x30s &   2018-08-06  & 58336.82467 & Blue & 25x60s   \\
            & 58215.91496      & Red & 20x30s  &   	       & 58336.80200  & Green & 25x60s  \\
2018-04-09  & 58217.98512  & Blue & 20x30s &   	       & 58336.77981  & Red & 25x60s    \\
            & 58217.99809   & Red & 20x30s  &   2018-08-13  & 58343.82051  & Blue & 25x60s   \\
2018-04-10  & 58218.00810   & Green & 20x30s&   	       & 58343.79584  & Green & 25x60s  \\
2018-04-28  & 58236.99101  & Blue & 10x60s &   	       & 58343.77401 & Red & 24x60s    \\
            & 58236.94531     & Green & 19x60s&   2018-08-22  & 58357.77011  & Green & 25x60s  \\
            & 58236.92411   & Red & 19x60s  &   	       & 58357.77011  & Red & 25x60s    \\
2018-05-03  & 58241.89642   & Blue & 20x60s &   2018-08-28  & 58358.78664  & Blue & 25x60s   \\
            & 58241.87706    & Green & 20x60s&   2018-09-12  & 58373.72876 & Blue & 25x60s   \\
            & 58241.85582   & Red & 20x60s  &   	       & 58373.75434  & Green & 25x60s  \\
2018-05-05  & 58243.91458   & Blue & 20x60s &   	       & 58373.78215 & Red &  25x60s   \\
            & 58243.89340    & Green & 20x60s&   2018-09-19  & 58378.73213 & Blue & 25x60s   \\
2018-05-10  & 58248.92518 & Blue & 20x60s &   	       & 58378.76053 & Green & 25x60s  \\
2018-05-20  & 58258.89443 & Blue & 20x60s &   	       & 58378.78497 & Red & 25x60s    \\
            & 58258.87651 & Green &20x60s &   2018-10-01  & 58392.70051  & Blue & 25x60s   \\
            & 58258.85870 & Red & 20x60s  &   	       & 58392.72588 & Green & 24x60s  \\
2018-05-24  & 58262.81056 & Green &20x60s &   	       & 58392.75174 & Red & 25x60s    \\
            & 58262.78823  & Red & 20x60s  &   2018-10-24  & 58415.66026    & Blue & 25x60s   \\
2018-05-30  & 58268.88980 & Green & 20x60s & &&&\\
            & 58268.86071 & Red & 20x60s  & &&&\\

\hline
\hline                             
\end{tabular}
\end{table*}

\section{Hardness-Intensity Diagrams}

\begin{center}
\begin{figure*}[!h]
\includegraphics[scale=0.45]{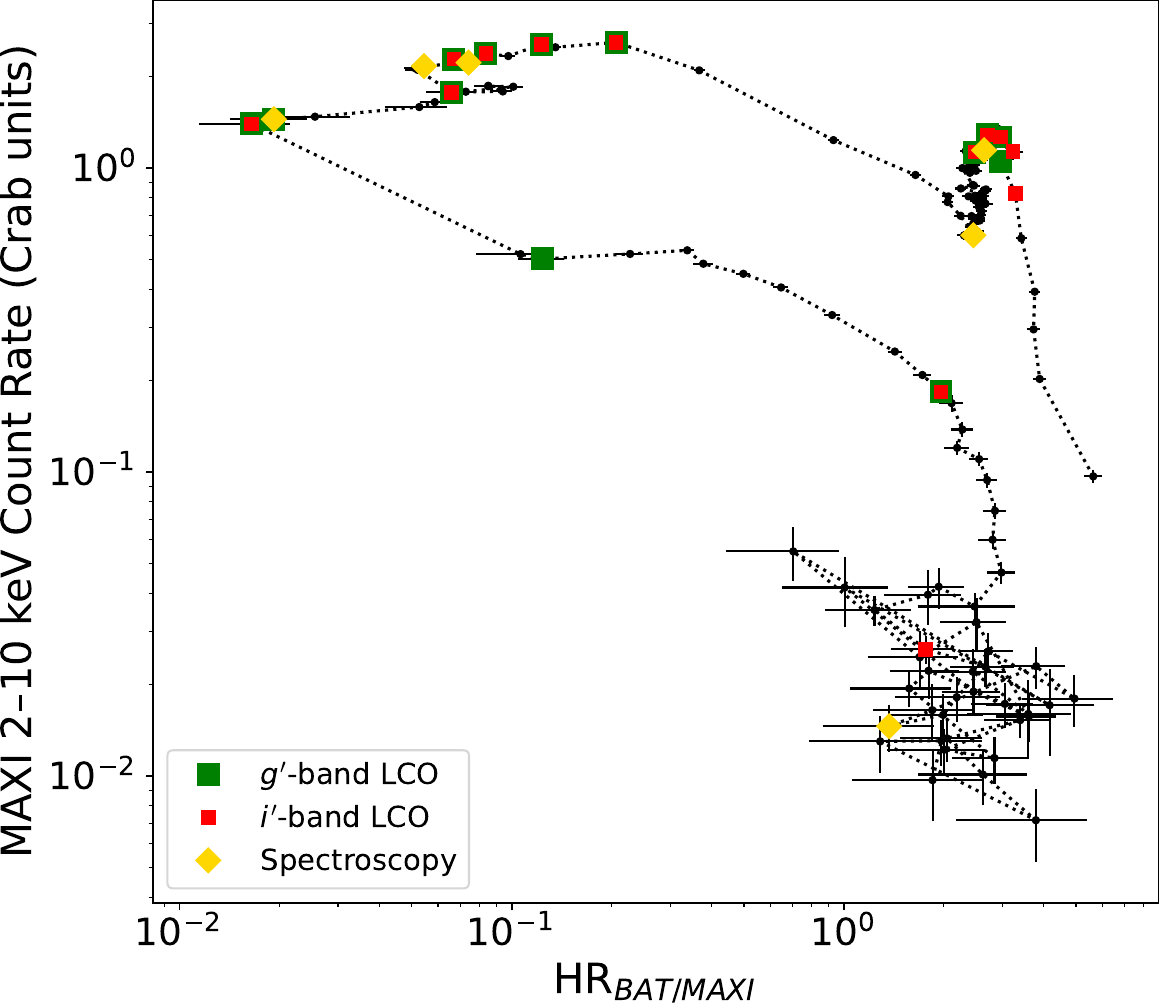}
\includegraphics[scale=0.45]{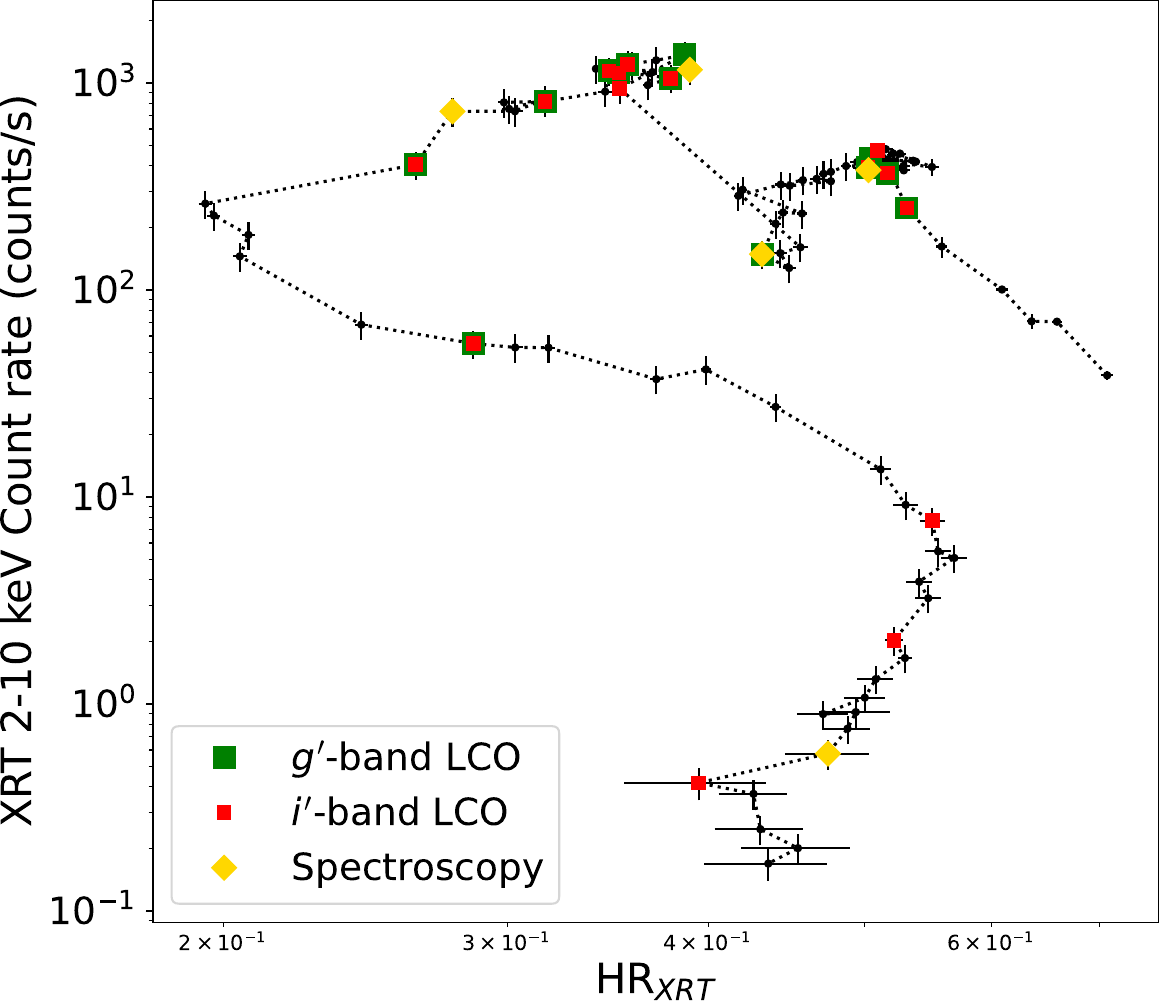}

\caption{Hardness-Intensity diagram of MAXI J1820+070 during its main 2018 outburst, constructed using $\rm HR_{BAT/MAXI}$ (\textit{left}) and $\rm HR_{XRT}$ (\textit{right}) as hardness ratios. Squares indicate epochs where nearly simultaneous (within one day) fractional rms variability amplitudes and hardness ratio estimates were obtained in the $g'$ (green) and $i'$ (red) LCO bands. For clarity, the Al Sadeem time-resolved epochs are not shown. Yellow diamonds mark the epochs of the spectroscopic campaign during the main outburst (Sec. \ref{spec} and Appendix \ref{app_spec}), with each point corresponding to the closest hardness ratio measurement to the spectroscopic observation. 
}
\label{fig_HID}
\end{figure*}
\end{center}

In order to observe the evolution of the hardness of the X-ray spectra, we built the hardness-intensity diagram \citep[HID;][]{Homan2001} of MAXI~J1820$+$070 during the main 2018 outburst, between MJD 58190 and MJD 58420 (that is, March 13 2018 and October 29, 2018, respectively). To do so, we used data acquired almost daily with \textit{Swift}/BAT and MAXI in the 15$-$50 keV and 2$-$10 keV energy bands, respectively. We converted the fluxes from both \textit{Swift}/BAT and MAXI into Crab units as described in Sec. \ref{sec:x-rays}, and we defined the hardness ratio $\rm HR_{\rm BAT/MAXI}$ as the ratio between the flux of \textit{Swift}/BAT and MAXI in Crab units. For the y-axis of the HID, we used the MAXI flux in Crab units in the 2$-$10 keV band (Fig. \ref{fig_HID}, left panel).  

Similarly, we built an HID using the hardness ratio $\rm HR_{\rm XRT}$, defined as the ratio of the 2$-$10 keV XRT count rates to the 0.5$-$2 keV XRT count rates, plotted against the XRT 2$-$10 keV count rates (Fig. \ref{fig_HID}, right panel). 

Following the spectral state classification for MAXI J1820+070 described by, e.g., \citet{Shidatsu2019} and \citet{Fabian2020}, the outburst displayed a standard pattern for a BH-LMXB: after an initial stage in the hard state, the system transitioned to a hard intermediate state around MJD 58300 (2018 July 1), rapidly transitioning to the soft state. The X-ray spectrum remained soft until $\sim$ MJD 58393 (2018 October 02), when it started to transition back to the hard state. After the transition to the hard state, the main outburst ended (as can be observed in the first two panels of Fig. \ref{lc_fig}). 




\section{Spectroscopic observations}\label{app_spec}
\begin{center}
\begin{figure*}[!h]
\includegraphics[scale=0.9]{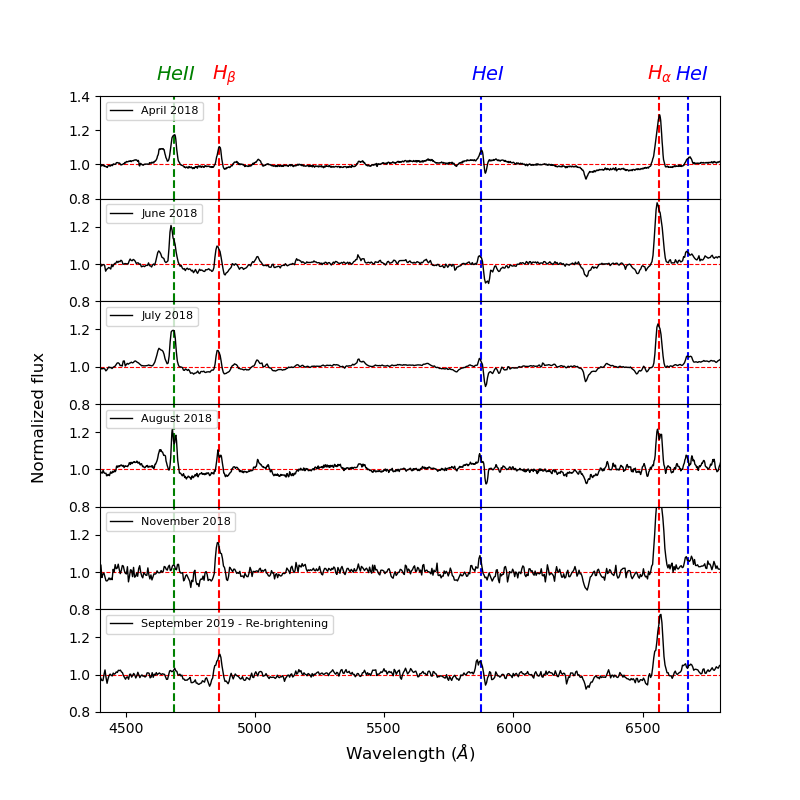}
\caption{Averaged spectra collected in the six different epochs reported in this work (Table \ref{tab:log_spec}). With dashed green, red and blue vertical lines, the position of the most prominent He II, H and He I emission lines is reported, respectively. A dashed red horizontal line marks where the normalized flux is equal to 1 in each panel.}
\label{fig:spectra}
\end{figure*}
\end{center}

The collection of spectra described in this work can be found in Fig. \ref{fig:spectra}. A complete log of the observations is reported in Tab. \ref{tab:log_spec}.
The spectra have been normalized to their local continua in order to facilitate a direct comparison of the line profiles. Although flux standards were observed during the runs, the resulting flux calibration is affected by significant and wavelength-dependent uncertainties (e.g. due to slit losses), and therefore the absolute flux level and continuum shape are not reliable. For this reason, we focus on the normalized spectra throughout the paper. In principle, the continuum shape could provide valuable information on the relative contributions of reprocessing and jet emission \citep[e.g.][]{Kimura2019}, as it may evolve from reprocessing-dominated to jet-dominated regimes. However, the spectral coverage and flux calibration accuracy of our data are insufficient for a reliable analysis of this kind.

\begin{table}[!h]
\caption{Complete log of the spectroscopic observations. The start time is indicated as YYYY-MM-DD HH:MM:SS.}            
\label{tab:log_spec}      
\centering                       
\begin{tabular}{c c c c}       
\hline    
Date Start time (UT)& MJD &Telescope & Exp \\
\hline
2018-04-15 10:12:20 & 58223.42523 & OAGH &1800 s\\
2018-04-15 10:42:29 & 58223.44617 & OAGH & 1800 s\\
\hline
2018-06-08 22:51:09 & 58277.95219 & Loiano & 900 s\\
2018-06-08 23:06:51 & 58277.96309 & Loiano & 900 s\\
\hline
2018-07-17 21:52:17 & 58316.91131 & Loiano & 1800 s\\
2018-07-18 19:59:59 & 58317.83332 & Loiano & 1800 s\\
\hline
2018-08-08 03:56:45 & 58338.16441 & OAN-SPM & 180 s\\
2018-08-08 04:00:07 & 58338.16675 & OAN-SPM & 180 s\\
2018-08-08 04:03:05 & 58338.16881 & OAN-SPM & 180 s\\
\hline
2018-11-09 17:41:41 & 58431.73728 & Loiano & 900 s\\
\hline
2019-09-20 18:52:09 & 58746.78621 & Loiano & 1800 s\\
2019-09-20 19:26:47 & 58746.81027 & Loiano & 1800 s\\
2019-09-20 20:01:43 & 58746.83452 & Loiano & 1800 s\\
\hline
\end{tabular}
\end{table}

\end{appendix}

\end{document}